# Spherical deconvolution of multichannel diffusion MRI data with non-Gaussian noise models and spatial regularization

Erick J. Canales-Rodríguez[1,2♬], Alessandro Daducci[3,4¶], Stamatios N. Sotiropoulos[5¶], Emmanuel Caruyer[6], Santiago Aja-Fernández[7], Joaquim Radua[1,2,8,9], Jesús M. Yurramendi Mendizabal[10], Yasser Iturria-Medina[11], Lester Melie-García[12], Yasser Alemán-Gómez[2,13,14], Jean-Philippe Thiran[3,4], Salvador Sarró[1,2], Edith Pomarol-Clotet[1,2], Raymond Salvador[1,2].

[1] FIDMAG Germanes Hospitalàries, C/ Dr. Antoni Pujadas, 38, 08830, Sant Boi de Llobregat, Barcelona, Spain. Tel: +34 93 6529999, Fax: +34 936400268.

[2] Centro de Investigación Biomédica en Red de Salud Mental, CIBERSAM, C/Dr Esquerdo, 46, 28007, Madrid, Spain.

[3] Signal Processing Lab (LTS5), École polytechnique fédérale de Lausanne (EPFL), Lausanne, Switzerland.

[4] University Hospital Center (CHUV) and University of Lausanne (UNIL), Lausanne, Switzerland.

[5] Centre for Functional Magnetic Resonance Imaging of the Brain (FMRIB), University of Oxford, John Radcliffe Hospital, Oxford OX39DU, United Kingdom.

[6] CNRS - IRISA (UMR 6074), Inria - VisAGeS Project-Team, INSERM - VisAGeS U746, Université de Rennes 1, Campus de Beaulieu, 35042 Rennes Cedex, France.

[7] Laboratorio de Procesado de Imagen (LPI), ETSI Telecomunicación, Universidad de Valladolid, Valladolid, Spain.

[8] Department of Psychosis Studies, Institute of Psychiatry, Psychology & Neuroscience, King's College London, United Kingdom.

[9] Department of Clinical Neuroscience, Karolinska Institutet, Stockholm, Sweden.

[10] Departamento de Ciencia de la Computación e Inteligencial Artificial, Universidad del País Vasco - Euskal Herriko Unibertsitatea, Spain.

[11] McConnell Brain Imaging Center, Montreal Neurological Institute, McGill University, Montreal, Quebec, Canada.

[12] The Neuroimaging Research Laboratory (Laboratoire de Recherche en Neuroimagerie: LREN), Department of Clinical Neurosciences, University Hospital Center (CHUV), Lausanne, Switzerland.

[13] Departamento de Bioingeniería e Ingeniería Aeroespacial. Universidad Carlos III de Madrid, Madrid, Spain.

[14] Instituto de Investigación Sanitaria Gregorio Marañón, Madrid, Spain.

♫ **Corresponding author**

Erick Jorge Canales-Rodríguez

E-mail: ejcanalesr@gmail.com

¶ These authors contributed equally to this work

# Abstract

Spherical deconvolution (SD) methods are widely used to estimate the intra-voxel white-matter fiber orientations from diffusion MRI data. However, while some of these methods assume a zero-mean Gaussian distribution for the underlying noise, its real distribution is known to be non-Gaussian and to depend on many factors such as the number of coils and the methodology used to combine multichannel MRI signals. Indeed, the two prevailing methods for multichannel signal combination lead to noise patterns better described by Rician and noncentral Chi distributions. Here we develop a Robust and Unbiased Model-BAsed Spherical Deconvolution (RUMBA-SD) technique, intended to deal with realistic MRI noise, based on a Richardson-Lucy (RL) algorithm adapted to Rician and noncentral Chi likelihood models. To quantify the benefits of using proper noise models, RUMBA-SD was compared with dRL-SD, a well-established method based on the RL algorithm for Gaussian noise. Another aim of the study was to quantify the impact of including a total variation (TV) spatial regularization term in the estimation framework. To do this, we developed TV spatially-regularized versions of both RUMBA-SD and dRL-SD algorithms. The evaluation was performed by comparing various quality metrics on 132 three-dimensional synthetic phantoms involving different inter-fiber angles and volume fractions, which were contaminated with noise mimicking patterns generated by data processing in multichannel scanners. The results demonstrate that the inclusion of proper likelihood models leads to an increased ability to resolve fiber crossings with smaller inter-fiber angles and to better detect non-dominant fibers. The inclusion of TV regularization dramatically improved the resolution power of both techniques. The above findings were also verified in human brain data.

# Introduction

After decades of developments in diffusion Magnetic Resonance Imaging (MRI), the successful implementation of a variety of advanced methods has shed light on the complex patterns of brain organization present at micro [1] and macroscopic scales [2-4]. Among these methods, Diffusion Tensor Imaging (DTI) [5] has become a classic in both clinical and research studies. DTI can deliver quantitative results, it may be easily implemented in any clinical MRI system and, thanks to its short acquisition time, it may be suitable for studying a wide range of brain diseases. Unfortunately, it is now well recognized that due to its simplistic assumptions, the DTI model does not adequately describe diffusion processes in areas of complex tissue organization, like in areas with kissing, branching or crossing fibers [6].

Such limitations in the DTI approach have prompted the recent development of numerous sampling protocols, diffusion models and reconstruction techniques (e.g., see [7-9] and references therein). While some of these techniques have been based on model-free methods, including q-ball imaging [10] and its extensions [11-17], diffusion orientation transforms [18, 19], diffusion spectrum imaging [20] and related q-space techniques [21-26], other approaches have relied on parametric diffusion models using higher-order tensors [27-29] and multiple second-order diffusion tensors [6]. In the later group, different numerical techniques involving gradient descent [6], Bayesian inference [30-32] and algorithms inspired from compressed sensing theory [33-35] have been applied to solve the resulting inverse problems.

Spherical Deconvolution (SD) is a class of multi-compartment reconstruction technique that can be implemented using both parametric and nonparametric signal models [36-49]. SD methods have become very popular owing to their ability to resolve fiber crossings with small inter-fiber angles in datasets acquired within a clinically feasible scan time. This resolving power is driven by the fact that, as opposed to model-free techniques that estimate the diffusion Orientational Distribution Function (ODF), the output from SD is directly the fiber ODF itself.

Among the different SD algorithms, Constrained Spherical Deconvolution (CSD) [39, 40] has been received with special interest due to its good performance and short computational time. In CSD, the average signal profile from white-matter regions of parallel fibers is first estimated, and afterwards, the fiber ODF is estimated by deconvolving the measured diffusion data in each voxel with this signal profile, which is also known as the single-fiber ‘response function’.

More recently and as an alternative to CSD, a new SD method based on a damped Richardson-Lucy algorithm adapted to Gaussian noise (dRL-SD) has been proposed [37, 42]. An extensive evaluation of both CSD and dRL-SD algorithms has revealed a superior ability to resolve low anisotropy crossing-fibers by CSD but a lower percentage of spurious fiber orientations and a lower over-all sensitivity to the selection of the response function by the dRL-SD approach [50]. This later feature is of great relevance since the assumption of a common response function for all brain tracts is a clear over-simplification of both methods, with the consequences of it minimized by the dRL-SD.

From an algorithmic perspective dRL-SD inherits the benefits of the standard RL deconvolution method applied with great success in diverse fields ranging from microscopy [51] to astronomy [52]. Remarkably, RL deconvolution is robust to the experimental noise and the obtained solution can be constrained to be non-negative without the need for including additional penalization functions in the estimation process. Moreover, from a modeling point of view, dRL-SD is implemented using an extended multi-compartment model that allows considering the partial volume effect in brain voxels with mixture of white matter (WM), gray matter (GM) and cerebrospinal fluid (CSF), a strategy that has been shown to be effective in reducing the occurrence of spurious fiber orientations [37].

However, in spite of the good properties of dRL-SD and other SD methods some methodological issues remain. These methods, to some extent, assume additivity and zero mean Gaussianity for the underlying noise and are potentially vulnerable to significant departures from such an assumption. Indeed, it is well known that the MRI noise is non-Gaussian and depends on many factors, including the number of coils in the scanner and the multichannel image combination method. Real experiments have shown that noise follows Rician [53] and noncentral Chi (nc-$\chi$) distributions [54] evidencing the inappropriateness of the Gaussian model. This issue is especially relevant in diffusion MRI data where the high b-values required to enhance the angular contrast lead to extremely low signal-to-noise (SNR) ratios. A recent study [55] has shown that different multichannel image combination methods can changes the properties of the signal and can have an effect on fiber orientation estimation.

On the other hand, the standard reconstruction in SD, based on a voxel-by-voxel fiber ODF estimation, although reasonable it may not be optimal in a global sense as it does not take into account the underlying spatial continuity of the image. Recent research on the inclusion of spatial continuity into SD methods via regularization has yielded promising results [9, 56, 57]**.** Among these, spatially regularized SD methods based on Total Variation (TV) [9] are very appealing due to their outstanding ability to simultaneously smooth away noise in flat regions whilst preserving edges, and due to their robustness to high levels of noise [58].

This work has two main aims: (1) the study and quantification of the benefits of the adequate modelling of the noise distribution in the context of spherical deconvolution, and (2) the study and quantification of the effects of including a TV spatial regularization term in the proposed estimation framework.

To address the first objective we developed a new SD methodology which, following a more realistic view, deals with non-Gaussian noise models. Specifically, the estimation framework is based on a natural extension to the RL algorithm for Rician and nc-$\chi$ likelihood distributions. We had chosen the RL algorithm as a starting point for our work because this algorithm has proven to be highly efficient in diverse applications, and because the performance of the resulting method can be directly compared to the state-of-the-art dRL-SD method, which employs a nearly-equivalent SD estimation algorithm but based on a Gaussian noise model. The second aim was addressed by including TV regularization to the developed formulation. Moreover, for completeness we have extended also the dRL-SD method via the spatially-regularized proposed estimation.

To compare the relative performance between the SD methods based on Gaussian and non-Gaussian noise models, and their respective implementations including the TV regularization, the different algorithms were applied to several synthetic phantom datasets which had been contaminated with noise patterns mimicking the Rician and nc-$\chi$ noise distributions produced in multichannel scanners. To the best of our knowledge, this is the first evaluation of such methods in a scenario where Rician and nc-$\chi$ noise are explicitly created as a function of the number of coils, their spatial sensitivity maps, the correlation between coils, and the reconstruction methodology used to combine the multichannel signals. As a final analysis, the new method is also applied to real multichannel diffusion MRI data from a healthy subject.

Following this introduction there is a 'Theory' section providing an overview of the different topics relevant to the study and the derivation of the new SD reconstruction algorithm. Description about computer simulations, image acquisition strategies and metrics designed to evaluate the performance of the reconstructions is provided in the Materials and Methods section. Relevant findings are succinctly described in the Results section. Finally, main results, contributions and limitations of this work are addressed in the Discussion and Conclusions section.

# Theory

This section contains a description of the forward/generative model used to relate the local diffusion process with the measured diffusion MRI data. It also provides a brief review of MRI noise models. Finally, the diffusion and MRI noise models are used to derive the new SD reconstruction algorithms.

## 1 Generative signal and fiber ODF model

The diffusion MRI signal measured for a given voxel can be expressed as the sum of the signals from each intra-voxel compartment. The term 'compartment' is defined as a homogeneous region in which the diffusion process possesses identical properties in magnitude and orientation throughout, and which is different to the diffusion processes occurring in other compartments. One example of this approach is the multi-tensor model that allows considering multiple WM parallel-fiber populations within the voxel. In this model the diffusion process taking place inside each compartment of parallel fibers is described by a second-order self-diffusion tensor [6].

In real brain data, in addition to the different WM compartments, voxels might also contain GM and CSF components. This issue was considered by [37], who extended the multi-tensor model by incorporating the possible contribution from these compartments. This is the generative multi-tissue signal model that will be used in the present study. In the absence of any source of noise, the resulting expression for the signal is:

$$S_i = S_0 \left( \sum_{j=1}^{M} \mathrm{f}_j \exp\left(-b_i \mathbf{v}_i^T \mathbf{D}_j \mathbf{v}_i\right) + \mathrm{f}_{GM} \exp\left(-b_i D_{GM}\right) + \mathrm{f}_{CSF} \exp\left(-b_i D_{CSF}\right) \right), \qquad \textbf{(1)}$$

where $M$ is the total number of WM parallel fiber bundles; $\mathrm{f}_j$ denotes the volume fraction of the $j$th fiber-bundle compartment; $\mathrm{f}_{GM}$ and $\mathrm{f}_{CSF}$ are the volume fractions of the GM and CSF compartments respectively, so that $\sum_{j=1}^{M} \text{f}_j + \text{f}_{GM} + \text{f}_{CSF} = 1$; $b_i$ is the diffusion-sensitization factor (i.e., $b$-value) used in the acquisition scheme to measure the diffusion signal $S_i$ along the diffusion-sensitizing gradient unit vector $\mathbf{v}_i$, $i = 1,...,N$; $D_{GM}$ and $D_{CSF}$ are respectively the mean diffusivity coefficients in GM and CSF; $S_0$ is the signal amplitude in the absence of diffusion-sensitization gradients ($b_i = 0$); $\mathbf{D}_j = \mathbf{R}_j^T \mathbf{A} \mathbf{R}_j$ denotes the anisotropic diffusion tensor of the $j$th fiber-bundle, where $\mathbf{R}_j$ is the rotation matrix that rotates a unit vector initially oriented along the x-axis toward the $j$th fiber orientation $(\theta_j, \phi_j)$ and $\mathbf{A}$ is a diagonal matrix containing information about the magnitude and anisotropy of the diffusion process inside that compartment:

$$\mathbf{A} = \begin{pmatrix} \lambda_1 & 0 & 0 \\ 0 & \lambda_2 & 0 \\ 0 & 0 & \lambda_3 \end{pmatrix}, \quad \textbf{(2)}$$

where $\lambda_1$ is the diffusivity along the $j$th fiber orientation, $\lambda_2$ and $\lambda_3$ are the diffusivities in the plane perpendicular to it. It is assumed that $\lambda_1 > \lambda_2 \approx \lambda_3$.

At each voxel, the measured diffusion signals $S_i$ for $N$ different sampling parameters (i.e., $\mathbf{v}_i$ and $b_i$, $i \in [1,...,N]$) can be recast in matrix form as:

$$\mathbf{S} = \mathbf{H}\mathbf{f}, \quad \textbf{(3)}$$

where $\mathbf{S} = [S_1 \quad \ldots \quad S_i \quad \ldots \quad S_N]^T$ and $\mathbf{H} = [\mathbf{H}^{WM} \quad | \quad \mathbf{H}^{ISO}]$ comprises two sub-matrices. $\mathbf{H}^{WM}$ is an $N$ x $M$ matrix where every column of length $N$ contains the values of the signal generated by the model given in Eq. (1) for a single fiber-bundle compartment oriented along one of the $M$-

directions, i.e., the $(i, j)$th element of $\mathbf{H}^{WM}$ is equal to $\mathbf{H}_{ij}^{WM} = S_0 \exp\left(-b_i \mathbf{v}_i^T \mathbf{D}_j \mathbf{v}_i\right)$. Likewise, $\mathbf{H}^{ISO}$ is an $N$ x $2$ matrix where each of the two columns of length $N$ contains the values of the signal for each isotropic compartment, i.e., $\mathbf{H}_{i1}^{ISO} = S_0 \exp\left(-b_i D_{GM}\right)$ and $\mathbf{H}_{i2}^{ISO} = S_0 \exp\left(-b_i D_{CSF}\right)$. Finally, the column-vector $\mathbf{f}$ of length $M+2$ includes the volume fractions of each compartment within the voxel.

In the framework of model-based spherical deconvolution, $\mathbf{H}$ is created by specifying the diffusivities, which are chosen according to prior information, and by providing a dense discrete set of equidistant $M$-orientations $\Omega = \left\{\left(\theta_j, \phi_j\right), j \in [1, \ldots, M]\right\}$ uniformly distributed on the unit sphere. Previous studies have used different sets of orientations, ranging from $M$=129 [43] to $M$=752 [42]. Then, the goal is to infer the volume fraction of all predefined oriented fibers, $\mathbf{f}$, from the vector of measurements $\mathbf{S}$ and the 'dictionary' $\mathbf{H}$ of oriented basis signals. Under this reconstruction model, $\mathbf{f}$ can be interpreted to as the fiber ODF evaluated on the set $\Omega$. Matrix $\mathbf{H}$ is also known as the 'diffusion basis functions' [43], or the 'point spread function' [37-39] that blurs the fiber ODF to produce the observed measurements.

It should be noticed that solving the deconvolution problem given by Eq. (3) is not simple because the resulting system of linear equations is ill-conditioned and ill-posed (i.e., there are more unknowns than measurements and some of the columns of $\mathbf{H}$ are highly correlated), which can lead to numerical instabilities and physically meaningless results (e.g., volume fractions with negative values). A common strategy to avoid such instabilities is to use robust algorithms that search for solutions compatible with the observed data but which also satisfy some additional constraints. Thus, in SD it is typical to estimate the fiber ODF by constraining it to be non-negative and symmetric around the origin (i.e., antipodal symmetry). As mentioned in the introduction, though, all these reconstruction algorithms may not be necessarily optimal when dealing with non-Gaussian noise, as it is the case for MRI noise.

# 2 MRI noise models

In conventional MRI systems, the data are measured using a single quadrature detector (i.e., coil with two orthogonal elements) that gives two signals which, for convenience, are treated as the real and imaginary parts of a complex number. The magnitude of this complex number (i.e. the square root of the sum of their squares) is commonly used because it avoids different kinds of MRI artifacts [53]. Given that the noise in the real and imaginary components follows a Gaussian distribution, the magnitude signal $S_i$ will follow a Rician distribution [53] with a probability function given by

$$P(S_i | \bar{S}_i, \sigma^2) = \frac{S_i}{\sigma^2} \exp\left\{ -\frac{1}{2\sigma^2} \left[ S_i^{\,2} + \bar{S}_i^2 \right] \right\} I_0\left( \frac{S_i \bar{S}_i}{\sigma^2} \right) u\left( S_i \right), \quad \textbf{(4)}$$

where $\bar{S}_i$ denotes the true magnitude signal intensity in the absence of noise, $\sigma^2$ is the variance of the Gaussian noise in the real and imaginary components, $I_0$ is the modified Bessel function of first kind of order zero and $u$ is the Heaviside step function that is equal to 0 for negative arguments and to 1 for non-negative arguments.

Modern clinical scanners are usually equipped with a set of 4 to 32 multiple phased-array coils, the signals of which can be combined following different strategies that, in turn, will give rise to different statistical properties for the noise [54]. One frequent strategy uses the spatial matched filter (SMF) approach linearly combining the complex signals of each coil and producing voxelwise complex signals [59]. Since the noise in the resulting real and imaginary components remains Gaussian a Rician distribution is expected in the final combined magnitude image. An alternative to the SMF is to create the composite magnitude image as the root of the sum-of-squares (SoS) of the complex signals of each coil. Under this approach the combined image follows a nc-χ distribution [60] given by,

$$P(S_i \left| \bar{S}_i, \sigma^2, n \right.) = \frac{\bar{S}_i}{\sigma^2} \left( \frac{S_i}{\bar{S}_i} \right)^n \exp\left\{ -\frac{1}{2\sigma^2} \left[ S_i^2 + \bar{S}_i^2 \right] \right\} I_{n-1}\left( \frac{S_i \bar{S}_i}{\sigma^2} \right) u\left( S_i \right), \tag{5}$$

where $n$ is the number of coils and $I_{n-1}$ is the modified Bessel function of first kind of order $n-1$. This expression is strictly valid when the different coils produce uncorrelated noise with equal variance, and when noise correlation cannot be neglected it provides a good approximation if effective $n_{eff}$ and $\sigma_{eff}^2$ values are considered [61], with $n_{eff}$ being a non-integer number lower than the real number of coils and $\sigma_{eff}^2$ is higher than the real noise variance in each coil.

A related SoS image combination method that increases the validity of Eq. (5) is the covariance-weighted SoS. This method is equivalent to pre-whitening (i.e., decorrelate) the measured signals before applying the standard SoS image combination. The covariance-weighted SoS approach requires the estimation of the noise covariance matrix of the system which, in practice, may be carried out by digitizing noise from the coils in the absence of excitations [62].

It is important to note that there are additional factors that can change the noise characteristics described above, including the use of accelerated techniques based on under-sampling approaches such as those used in parallel MRI (pMRI) and partial Fourier, certain reconstruction filters in k-space, and some of the preprocessing steps conducted after image reconstruction.

Empirical data suggest that some of these factors do not substantially change the type of distribution of the noise. On the one hand, [54] investigated the effects of the type of filter in k-space, the number of receiving channels and the use of pMRI reconstruction techniques, and found that noise distributions always followed Rician and nc-$\chi$ distributions with a reasonable accuracy - although their standard deviations and effective number of receiver channels were altered when fast pMRI and subsequent SoS reconstructions were used. On the other hand, [55] showed real diffusion MRI data noise to also follow Rician and nc-$\chi$ noise distributions after a preprocessing that included motion and eddy currents corrections. Unfortunately, the combined effect of all factors has not, to the best of our knowledge, being studied. In this regard, a complete evaluation should include the study of the effects of additional data manipulation processes routinely applied in many clinical research studies, such as B0-unwarping due to magnetic field inhomogeneity and partial Fourier reconstructions. Although the latter has been investigated in terms of signal-to-noise ratio, its influence on the shape of the noise distribution remains unknown.

However, while it is impossible to ensure that Rician and nc-χ distributions are the optimal noise models for all possible strategies used for sampling, reconstructing and preprocessing diffusion MRI data, such models are flexible enough to adapt to deviations from the initial theoretical assumptions. Their parameterization in terms of spatial-dependent effective parameters (i.e., $n_{eff}\left(x,y,z\right)$, $\sigma_{eff}^{2}\left(x,y,z\right)$, as in [61, 63, 64]) allows characterizing the spatially varying nature of the noise observed in accelerated MRI reconstructed data, as well as the spatial correlation introduced by reconstruction algorithms, whilst preserving the good theoretical properties of the models with standard parameters, i.e. the null probability to obtain negative signals and the ability to characterize the signal-dependent non-linear bias of the data.

# 3 Spherical deconvolution of diffusion MRI data

Equation (5) based on either conventional (i.e., $n$, $\sigma^2$) or effective (i.e., $n_{eff}$, $\sigma_{eff}^{2}$) parameters provides a very general MRI noise model, which includes the Rician distribution (given in Eq. (4)) as a special case with $n=1$. Consequently, if we derive the spherical deconvolution reconstruction corresponding to Equation (6) any particular solution of interest will become available.

Specifically, if we assume the linear model given by Eqs. (1)-(3) the likelihood model for the vector of measurements $\mathbf{S}$ under a nc-χ distribution is

$$P(\mathbf{S}|\mathbf{H},\mathbf{f},\sigma^2,n)=\prod_{i=1}^{N}\frac{\bar{S}_i}{\sigma^2}\left(\frac{S_i}{\bar{S}_i}\right)^n \exp\left\{-\frac{1}{2\sigma^2}\left[S_i^2+\bar{S}_i^2\right]\right\} I_{n-1}\left(\frac{S_i\bar{S}_i}{\sigma^2}\right)u\left(S_i\right), \quad \textbf{(6)}$$

where $S_i$ and $\bar{S}_i = (\mathbf{Hf})_i$ are the measured and expected signal intensities for $i$th sampling parameters, respectively.

## 3.1 Unbiased and positive recovery: the multiplicative Richardson-Lucy algorithm for nc-χ noise

The maximum likelihood (ML) estimate in Eq. (6) is obtained by differentiating its negative log-likelihood $J(\mathbf{f}) = -\log P(\mathbf{S}|\mathbf{H},\mathbf{f},\sigma^2,n)$ with respect to $\mathbf{f}$ and equating the derivative to zero, which after some algebraic manipulations becomes

$$\mathbf{f} = \mathbf{f} \circ \frac{\mathbf{H}^T \left[ \mathbf{S} \circ \dfrac{I_n\left(\mathbf{S} \circ \mathbf{Hf}/\sigma^2\right)}{I_{n-1}\left(\mathbf{S} \circ \mathbf{Hf}/\sigma^2\right)} \right]}{\mathbf{H}^T\mathbf{Hf}}, \qquad \textbf{(7)}$$

where '$\circ$' stands for the Hadamard component-wise product, and the division operators are applied component-wise to the vector's elements.

Equation (7) is nonlinear in $\mathbf{f}$ and its solution can be obtained through a modified version of the expectation maximization technique, originally developed by Richardson and Lucy for a Poisson noise model [65, 66] and known as the RL algorithm. When we applied this technique to nc-χ and Rician distributed noise it naturally led to the following iterative estimation formula:

$$\mathbf{f}^{k+1} = \mathbf{f}^k \circ \frac{\mathbf{H}^T \left[ \mathbf{S} \circ \dfrac{I_n\left(\mathbf{S} \circ \mathbf{Hf}^k/\sigma^2\right)}{I_{n-1}\left(\mathbf{S} \circ \mathbf{Hf}^k/\sigma^2\right)} \right]}{\mathbf{H}^T\mathbf{Hf}^k}, \qquad (8)$$

in which the solution calculated at the $k$th iteration step ($\mathbf{f}^k$) gradually improves (i.e. its likelihood increases after each step) until a final, stationary solution $\frac{\mathbf{f}^{k+1}}{\mathbf{f}^k}=1$, is reached. As shown Appendix A in S1 File, this formula can also be related to the RL algorithm for Gaussian noise, employed in the undamped RL-SD technique [42].

Under the absence of any prior knowledge about $\mathbf{f}$, the initial estimate ($\mathbf{f}^0$) can be fixed to a non-negative constant density distribution [42]. In that case, the algorithm transforms a perfectly smooth initial estimate into sharper estimates, with sharpness increasing with the number of iterations. So, roughly speaking, the number of iterations can be considered as a regularization parameter controlling the angular smoothness of the final estimate. Notably, if $\mathbf{f}^0$ is non-negative, the successive estimates remain non-negative as well, and the algorithm always produces reconstructions with positive elements. Moreover, as in [37, 42] the estimation does not involve any matrix inversion, thus avoiding related numerical instabilities.

In order to evaluate Eq. (8) an estimate $\tilde{\sigma}^2$ of $\sigma^2$ is required. Although obtaining it from a region-of-interest (ROI) is feasible [67] its accuracy may be compromised by systematic experimental issues such as ghosting artifacts, signal suppression by the scanner outside the brain, zero-padding and by filters applied in the *k*-space. Moreover, with the use of fast parallel MRI sequences, where each coil records signals with partial coverage in the k-space, properties of the noise become spatially heterogeneous (i.e. they change from voxel to voxel across the image). While some authors have proposed alternatives to overcome these limitations [68] here we have estimated the noise variance at each voxel from the same data used to infer the fiber ODF.

Specifically, by minimizing the negative log-likelihood with respect to $\sigma^2$ we have obtained an iterative scheme analogous to Eq. (8):

$$\alpha^{k+1}=\frac{1}{nN}\left\{\frac{\mathbf{S}^T\mathbf{S}+\mathbf{f}^T\mathbf{H}^T\mathbf{H}\mathbf{f}}{2}-\mathbf{1}_N^T\left[\left(\mathbf{S}\circ\mathbf{H}\mathbf{f}\right)\circ\frac{I_n\left(\mathbf{S}\circ\mathbf{H}\mathbf{f}/\alpha^k\right)}{I_{n-1}\left(\mathbf{S}\circ\mathbf{H}\mathbf{f}/\alpha^k\right)}\right]\right\}, \quad \mathbf{(9)}$$

where $\alpha^k$ is the estimate of $\sigma^2$ at the $k$ th iteration (starting with an arbitrarily initial estimate $\alpha^0$) and $\mathbf{1}_N$ is a $N \times 1$ vector of ones. The resulting algorithm based on equations (8) and (9) is termed RUMBA-SD, which is the abbreviation of '**R**obust and **U**nbiased **M**odel-**Ba**sed **S**pherical **D**econvolution'. The spatially-regularized extension to this algorithm is described in the following section.

## 3.2 Towards a robust recovery: Total variation regularization

When considering the TV model [58] the maximum a posteriori (MAP) solution at voxel $(x, y, z)$ is obtained by minimizing the augmented functional:

$$J(\mathbf{f}) = -\log P(\mathbf{S} | \mathbf{H}, \mathbf{f}, \sigma^2, n) + \alpha_{TV} TV(\mathbf{f}), \quad \textbf{(10)}$$

where the first term is the negative log-likelihood defined in previous sections and the second term is the TV energy, defined as the sum of the absolute values of the first-order spatial derivative (i.e., gradient "$\nabla$") of the fiber ODF components over the entire brain image, $TV(\mathbf{f}) = \sum_j \left| \nabla [\mathbf{f}_{3D}]_j \right|$, evaluated at voxel $(x, y, z)$; $[\mathbf{f}_{3D}]_j$ is a 3D image created in a way that each voxel contains the element at position $j$ of their corresponding estimate vector $\mathbf{f}$, and $\alpha_{TV}$ is a parameter controlling the level of spatial regularization. Importantly, and in contrast to the previous ML estimate, now the solution at a given voxel is not independent from the solutions in other voxels. The spatial dependence introduced by the TV functional promotes smooth solutions in homogeneous regions (discourages the solution from having oscillations), yet it does allow the solution to have sharp discontinuities [58]. This property is highly relevant for SD because, while it promotes continuity and smoothness along individual tracts, it prevents partial volume contamination from adjacent tracts.

In this work, the MAP estimate from Eq.(10) is obtained using an iterative scheme similar to that proposed in [51], where the estimate at each iteration is calculated by the multiplication of two terms: the standard ML estimate, and the regularization term derived from the TV functional

$$\mathbf{f}^{k+1} = \mathbf{f}^{k} \circ \frac{\mathbf{H}^{T}\left[\mathbf{S} \circ \dfrac{I_{n}\left(\mathbf{S} \circ \mathbf{H}\mathbf{f}^{k}/\sigma^{2}\right)}{I_{n-1}\left(\mathbf{S} \circ \mathbf{H}\mathbf{f}^{k}/\sigma^{2}\right)}\right]}{\mathbf{H}^{T}\mathbf{H}\mathbf{f}^{k}} \circ \mathbf{R}^{k}, \quad \textbf{(11)}$$

with the TV regularization vector $\mathbf{R}^{k}$ at voxel $(x, y, z)$, and at the $k$th iteration, computed element-by-element as

$$\left(\mathbf{R}^{k}\right)_{j} = \frac{1}{1 - \alpha_{TV}\, div\left(\dfrac{\nabla\left[\mathbf{f}_{3D}^{k}\right]_{j}}{\left|\nabla\left[\mathbf{f}_{3D}^{k}\right]_{j}\right|}\right)\Bigg|_{(x,y,z)}}, \quad \textbf{(12)}$$

where $\left(\mathbf{R}^{k}\right)_{j}$ is the element $j$ of vector $\mathbf{R}^{k}$ and $div$ is the divergence operator. In practice, to correct for potential singularities at $\left|\nabla\left[\mathbf{f}_{3D}^{k}\right]_{j}\right| = 0$, the term $\left|\nabla\left[\mathbf{f}_{3D}^{k}\right]_{j}\right|$ is replaced by its approximated value $\sqrt{\left|\nabla\left[\mathbf{f}_{3D}^{k}\right]_{j}\right|^{2} + \varepsilon}$, where $\varepsilon$ is a small positive constant. Moreover, any negative value in $\mathbf{R}^{k}$ is replaced by its absolute value to preserve the non-negativity of the estimated fiber ODF. Notice that by setting $\alpha_{TV} = 0$ the estimator in Eq. (11) becomes equal to the unregularized version in Eq. (8).

In the current implementation the simultaneous estimation of all the parameters is carried out via an alternating iterative scheme summarized in Table 1. Briefly, it minimizes the functional in Eq.

(10) with respect to the fiber ODF while assuming that $\sigma^2$ is known and fixed, and then it updates the noise variance using the new fiber ODF estimate. While for SMF-based data all the equations are evaluated using $n=1$, for SoS-based data $n$ is fixed to the real number of coils, or to the effective value $n_{eff}$ if provided. (But see Appendix B in S1 File)

**Table 1**. General pseudocode MAP algorithm.

| |
|---|
| Initialize $\mathbf{f}^0$ and $\alpha^0$ |
| *if* SMF, then |
| $n = 1$ |
| *else if* SOS, then |
| $n$ = number of coils (or $n_{eff}$) |
| *end* |
| *for* $k$ = 1,2,…, repeat the following steps, until a termination criterion is satisfied |
| compute $\mathbf{f}^{k+1}$ via Eqs. (11) and (12), assuming $\sigma^2 = \alpha^k$ |
| $\mathbf{f}^{k+1} = \mathbf{f}^{k+1}$ /sum($\mathbf{f}^{k+1}$) (*) |
| compute $\alpha^{k+1}$ via Eq. (9) assuming $\mathbf{f} = \mathbf{f}^{k+1}$ |
| update $\alpha_{TV}$ |
| *end* |

(*) Optionally, the ODF vector may be scaled to unity, thus preserving the physical definition of the $j$ th element in $\mathbf{f}$ as the volume fraction of the $j$ th compartment of the voxel (see Eqs. (1)-(3)). This step would make sense when the fiber response signal used to create the dictionary matches the real signal from the compartments, whereas it may be omitted when the latter cannot be guaranteed. Notice that the original implementation of dRL-SD did not include this step.

The regularization parameter $\alpha_{TV}$ is adaptively adjusted at each iteration following the discrepancy principle. Specifically, it is selected to match the estimated variance [69] using two alternative strategies: (i) assuming a constant mean parameter over the entire brain image, $\alpha_{TV} = E\{\alpha^{k+1}\}$ (see Table 1), potentially increasing the precision and robustness of the estimator; or (ii) assuming a spatial dependent parameter, $\alpha_{TV}(x, y, z) = \alpha^{k+1}(x, y, z)$, which may be more appropriate in situations where a differential variance across the image is expected, like in accelerated pMRI based-data.

It should be remembered that the accuracy of the reconstruction for the SoS case depends on the variant used to combine the images. In this work it is assumed that the data is combined using the covariance-weighted SoS method. However, even if the available data were combined using the conventional SoS approach (i.e., without taking into account the noise correlation matrix among coil elements), the method could still provide a reasonable approximation (for more details see Appendix B in S1 File).

The evaluation of the ratio of modified Bessel functions of first kind involved in the updates of Eqs. (11) and (9) is best computed by considering the ratio as a new composite function, and not by means of the simple evaluation of the ratio of the individual functions. Specifically, this ratio is computed here in terms of Perron continued fraction [70]. All the details are provided in Appendix C in S1 File.

Following a similar estimation framework, the TV regularization was also included into the dRL-SD method. All the relevant equations are provided in Appendix D in S1 File.

# Materials and Methods

## 1 Synthetic fiber bundles with different inter-fiber angles

In order to test the resolving power of the methods as a function of the underlying inter-fiber angle, various synthetic phantoms including two fiber bundles were generated. The inter-fiber angles were gradually modified from 1 to 90 degrees applying one degree increases, eventually yielding 90 different phantoms with 50 x 50 x 50 voxels. The volume fractions of the two fiber bundles were assumed to be equal ($f_1 = f_2 = 0.5$) in the fiber crossing region.

The intra-voxel diffusion MRI signal was generated via the multi-tensor model [6] using $N$ = 70 sampling orientations with constant $b$=3000 s/mm$^2$ plus one additional image with $b$=0 (i.e., $S_0 = 1$ was assumed in all voxels). The diffusion tensor diffusivities of both fiber groups were assumed to be identical and equal to $\lambda_1 = 1.7$ $10^{-3}$ mm$^2$/s and $\lambda_2 = \lambda_3 = 0.3$ $10^{-3}$ mm$^2$/s respectively.

## 2 Synthetic fiber bundles with different volume fractions

To test the ability of the different methods to detect non-dominant fibers, various synthetic phantoms containing two fiber bundles were generated. In these phantoms the inter-fiber angle was fixed to 70 degrees (an angle presumably detectable by all the methods) and the volume fraction of the non-dominant fiber bundle was gradually changed from 0.1 to 0.5 in 0.01steps, generating 41 different phantoms with 50 x 50 x 50 voxels each. The intra-voxel diffusion MRI signal was created using the same generative multi-tensor model, $b$-value and sampling orientations as in the previous section.

# 3 Synthetic "HARDI Reconstruction Challenge 2013" phantom

The reconstruction algorithms were also tested on the synthetic diffusion MRI phantom developed for the "HARDI Reconstruction Challenge 2013" Workshop, within the IEEE International Symposium on Biomedical Imaging (ISBI 2013). This phantom comprises a set of 27 fiber bundles with fibers of varying radii and geometry which connect different areas of a 3D image with 50 x 50 x 50 voxels. It contains a wide range of configurations including branching, crossing and kissing fibers, together with the presence of isotropic compartments mimicking the CSF contamination effects occurring near ventricles in real brain images.

The intra-voxel diffusion MRI signal was generated using $N$ = 64 sampling points on a sphere in q-space with constant $b$=3000 s/mm$^2$, plus one additional image with $b$=0. For pure GM and CSF voxels, signals were generated using two mono-exponential models: $\exp(-D_{GM}b)$ and $\exp(-D_{CSF}b)$ with $D_{GM}$ = 0.2 10$^{-3}$ mm$^2$/s and $D_{CSF}$ = 1.7 10$^{-3}$ mm$^2$/s. In voxels belonging to single-fiber WM bundles, the signal measured along the q-space unit direction $\hat{\mathbf{q}} = \mathbf{q}/|\mathbf{q}|$ was generated by a mixture of signals from intra- and extra-axonal compartments: $\mathrm{f}_{\mathrm{int}}\mathrm{s}_{\mathrm{int}}(\mathbf{q}, \mathbf{v}, \tau, L, R) + \mathrm{f}_{\mathrm{ext}}\mathrm{s}_{\mathrm{ext}}(\hat{\mathbf{q}}, \mathbf{v}, b, \lambda_1, \lambda_2)$, where $\mathbf{v}$ denotes the local fiber orientation. Signal from the intra-axonal compartment $s_{\mathrm{int}}$ was created following the theoretical model of a restricted diffusion process inside a cylinder of length $L = 5\ mm$ and radius $R = 5\ \mu m$ at the diffusion time $\tau$ =20.8 $s$ [19, 71]. The extra-axonal signal $\mathrm{s}_{\mathrm{ext}}$ was generated using a diffusion tensor model with cylindrical symmetry (i.e., $\lambda_1 = 1.7$ 10$^{-3}$ mm$^2$/s, $\lambda_2 = \lambda_3 = 0.2$ 10$^{-3}$ mm$^2$/s). Mixture fractions were fixed to $\mathrm{f}_{\mathrm{int}}$ = 0.6 and $\mathrm{f}_{\mathrm{ext}}$ = 0.4. The noiseless dataset can be freely downloaded from the Webpage of the site: http://hardi.epfl.ch/static/events/2013_ISBI/.

# 4 Multichannel noise generation

The synthetic diffusion images from the above phantoms were contaminated with noise mimicking the SoS and SMF strategies used in scanners in order to combine multiple-coils signals. To that aim, the noisy complex-valued image measured from the $k$th coil was assumed to be equal to

$$S_k = SC_k + e_k^R + ie_k^I, \qquad \textbf{(13)}$$

where $C_k$ is the relative sensitivity map [72] of $k$th coil, $e_k^R \sim N(0,\Sigma)$ and $e_k^I \sim N(0,\Sigma)$ are two different Gaussian noise realizations simulating the noise in the real and imaginary components with zero-mean value and covariance matrix $\Sigma$. For simplicity $\Sigma$ was assumed to be given by

$$\Sigma = \sigma^2 \begin{pmatrix} 1 & \rho & \cdots & \rho \\ \rho & 1 & \cdots & \rho \\ \vdots & \vdots & \ddots & \vdots \\ \rho & \rho & \cdots & 1 \end{pmatrix}, \qquad \textbf{(14)}$$

where $\sigma^2$ is the noise variance of each coil and $\rho$ indicates the correlation coefficient between any two coils.

For the SoS reconstruction, magnitude images were generated as:

$$S_{SoS} = \sqrt{\sum_{k=1}^{n} |S_k|^2}, \qquad \textbf{(15)}$$

where $|S_k|$ stands for the magnitude of $S_k$. Notice that Eq. (15) is the conventional SoS image combination and not the covariance-weighted variant. We have followed this approach in order to simulate the effect of any remaining residual correlation $\rho$ present in real systems (we have assumed a $\rho$ = 0.05) [63].

In the SMF reconstruction, magnitude images were generated as:

$$S_{SMF} = \left| \sum_{k=1}^{n} S_k C_k \right|, \quad \textbf{(16)}$$

with simulated sensitivity maps depicted in Fig A in S2 File satisfying the relationship $\sum_{k=1}^{n} C_k^2 = 1$, which holds in practice when the relative sensitivity maps are calculated as $C_k = |S_k| / S_{SoS}$ [72]. These sensitivity maps have been previously used in [73].

It should be noted that different scanner vendors can implement different SMF and SoS variants. In this work we have used the variants given in [55] for datasets acquired without undersampling in the k-space, i.e., R=1, where R is the acceleration factor of the acquisition defined as the ratio of the total k-space phase-encoding lines over the number of k-space lines actually acquired. Notice that in the absence of noise $S_{SoS} = S_{SMF}$. Besides, for the particular case of a single coil with uniform sensitivity, i.e., $n = 1$ and $C = 1$, Eq. (15) and Eq. (16) become identical.

The 132 3D phantoms resulting from the procedures described in the three previous sections were contaminated with noise using a range of clinical signal-to-noise ratios (SNR) of 10, 15, 20 and 30, where $SNR = S_0 / \sigma$. In order to generate signals under equivalent conditions, for each value of $\sigma$ the same noise realizations $\{e_k^R\}$ and $\{e_k^I\}$ were used to generate the final images $S_{SoS}$ and $S_{SMF}$. All datasets were created simulating a scanner with 8 coils (see Figs A and B in S2 File).

# 5 Evaluation metrics

The performance of the algorithms was quantified by comparing the obtained reconstructions against the ground-truth via three main criteria: (i) the angular error in the orientation of fiber populations, (ii) the proper estimation of the number of fiber populations present in every voxel and (iii) the volume fraction error.

For the analyses, local peaks from the reconstructed fiber ODFs were identified as those vertices in the grid with higher values than their adjacent neighbors, considering only cases where magnitudes exceeded at least one tenth of the amplitude of the highest peak (i.e., $0.1 \cdot f_{max}$) [50]. From all identified peaks, the highest four where finally retained.

Next, we adopted some of the evaluation metrics widely used in the literature [9]. Specifically, we used the angular error, defined as the average minimum angle between the extracted peaks and the true fiber directions [74]:

$$\theta = \frac{1}{M_{true}} \sum_{k=1}^{M_{true}} \min_m \left\{ \arccos\left( \left| \mathbf{e}_m^T \mathbf{v}_k \right| \right) \right\}, \qquad \textbf{(17)}$$

where $M_{true}$ is the true number of fiber populations, $\mathbf{e}_m$ is the unitary vector along with the $m$th detected fiber peak and $\mathbf{v}_k$ is the unitary vector along the $k$th true fiber direction. The volume fraction error of the estimated fiber compartments was assessed by means of the average absolute error between the estimated and the actual peak amplitudes:

$$\Delta f = \frac{1}{M_{true}} \sum_{k=1}^{M_{true}} \left| f_m - f_k \right|, \qquad \textbf{(18)}$$

where $f_m$ is the normalized height of the $m$th detected fiber peak and $f_k$ is the volume fraction of the $k$th true fiber. As usual, the angular and volume fraction errors between each pair of fibers were measured by comparing the true fiber with the closest estimated fiber.

Finally, the success rate (SR) was employed to quantify the estimation of the number of fiber compartments. The SR is defined as the proportion of voxels in which the algorithms estimate the right number of fiber compartments. To discriminate the different factors leading to an erroneous estimation, the mean number of over-estimated $n^+$ and under-estimated $n^-$ fiber populations were computed over the whole image [9].

# 6 Settings for the evaluation algorithms

Both RUMBA-SD and dRL-SD methods were implemented using in-house developed *Matlab* code, applying the same dictionaries **H** created from the signal generative model given in Eqs.(1)-(3). These used M = 724 fiber orientations distributed on the unit sphere, with a mean angular separation between adjacent neighbor vertices of 8.36 degrees, and a standard deviation of 1.18 degrees.

To assess the effect of using dictionaries with optimal and non-optimal diffusivities, two different dictionaries were created and applied to the datasets described in subsections 1 (i.e., fiber bundles with different inter-fiber angles) and 2 (i.e., fiber bundles with different volume fractions). The first dictionary was generated by using the same diffusivities employed in the synthetic data, whereas the second dictionary was created from diffusivities estimated in regions of parallel fibers (outside the fiber crossing area) by means of a standard diffusion tensor fitting on the noisy data (i.e., *dtifit* tool in FSL package).

Similarly, two dictionaries were created to test the reconstructions on the data described in subsection 3 (i.e., "HARDI Reconstruction Challenge 2013" phantom data). In this case the model diffusivities and the 'true' diffusivities were deliberately set to different values in order to consider the possibility of model misspecification. The first dictionary was created with tensor diffusivities equal to $\lambda_1 = 1.4\ 10^{-3}$ mm$^2$/s and $\lambda_2 = \lambda_3 = 0.4\ 10^{-3}$ mm$^2$/s. Two isotropic compartments with diffusivities equal to 0.2 $10^{-3}$ mm$^2$/s and 1.4 $10^{-3}$ mm$^2$/s were also included. In the second dictionary the diffusivities were assumed to be equal to $\lambda_1 = 1.6\ 10^{-3}$ mm$^2$/s and $\lambda_2 = \lambda_3 = 0.3\ 10^{-3}$ mm$^2$/s, and the isotropic diffusivities were equal to 0.2 $10^{-3}$ mm$^2$/s and 1.6 $10^{-3}$ mm$^2$/s respectively.

The starting condition $\mathbf{f}^0$ in all cases was set as a non-negative iso-probable spherical function [37]. The accuracy and convergence of both methods as a function of the number of iterations was investigated by repeating the calculations using 200 and 400 iterations, which is within the optimal range as suggested in [37] and [50]. The extended algorithms with TV regularization were also tested using 600 and 1000 iterations. The geometric damping and threshold parameters for dRL-SD were set to $\nu$=8 and $\eta$=0.06 respectively [37], see Appendix D in S1 File. For SoS-based data, $n$ was fixed to the real number of coils in RUMBA-SD.

To differentiate the standard RUMBA-SD and dRL-SD algorithms from their regularized versions, we have added the term '+TV' to their names, i.e., RUMBA-SD+TV and dRL-SD+TV.

# 7 Real brain data

Diffusion MRI data were acquired from a healthy subject on a 3T Siemens scanner (Erlangen) located at the University of Oxford (UK). The subject provided informed written consent before participating in the study, which was approved by the Institutional Review Board of the University of Oxford. Whole brain diffusion images were acquired with a 32-channel head coil along 256 different gradient directions on the sphere in q-space with constant $b$ = 2500 s/mm$^2$. Additionally, 36 $b$ = 0 volumes were acquired with in-plane resolution = 2.0 x 2.0 mm$^2$ and slice thickness = 2 mm. The acquisition was carried out without undersampling in the $k$-space (i.e., R=1). Raw multichannel signals were combined using either the standard GRAPPA approach or the GRAPPA approach with the adaptive combination of the SMF available in the scanner, giving SoS and SMF-based datasets respectively. Then, the two resulting datasets were separately corrected for eddy current distortions and head motion as implemented in FSL [75].

A subset of 64 directions with nearly uniform coverage on the sphere was selected from the full set of 256 gradients directions, and measurements for this subset were used to 'create' an under-sampled version of the data, which also included 3 $b$=0 volumes. The resulting HARDI sequence based on 64 directions is similar to those widely employed in clinical studies, thus results from this dataset are useful to evaluate the impact of the new technique on standard clinical data.

# Results

## 1 Gaussian versus non-Gaussian noise models

The angular and volume fraction errors from the dRL-SD and RUMBA-SD reconstructions in the 90 synthetic phantoms with different inter-fiber angles are depicted in Fig 1, as well as in Fig C in S2 File. Fig 1 shows results using a dictionary created with the same diffusivities applied to generate the data (i.e., $\lambda_1 = 1.7\ 10^{-3}$ mm$^2$/s and $\lambda_2 = \lambda_3 = 0.3\ 10^{-3}$ mm$^2$/s), whereas Fig C in S2 File displays results using tensor diffusivities estimated from the noisy data ($\lambda_1 = 1.1\ 10^{-3}$ mm$^2$/s and $\lambda_2 = \lambda_3 = 0.35\ 10^{-3}$ mm$^2$/s). In both dictionaries two isotropic compartments with diffusivities equal to 0.1 $10^{-3}$ mm$^2$/s and 2.5 $10^{-3}$ mm$^2$/s were included. Average values of SR, $n^+$ and $n^-$ are also reported in Figs D and E in S2 File. Results shown come from the reconstructions employing 200 iterations and the datasets with a SNR=15.

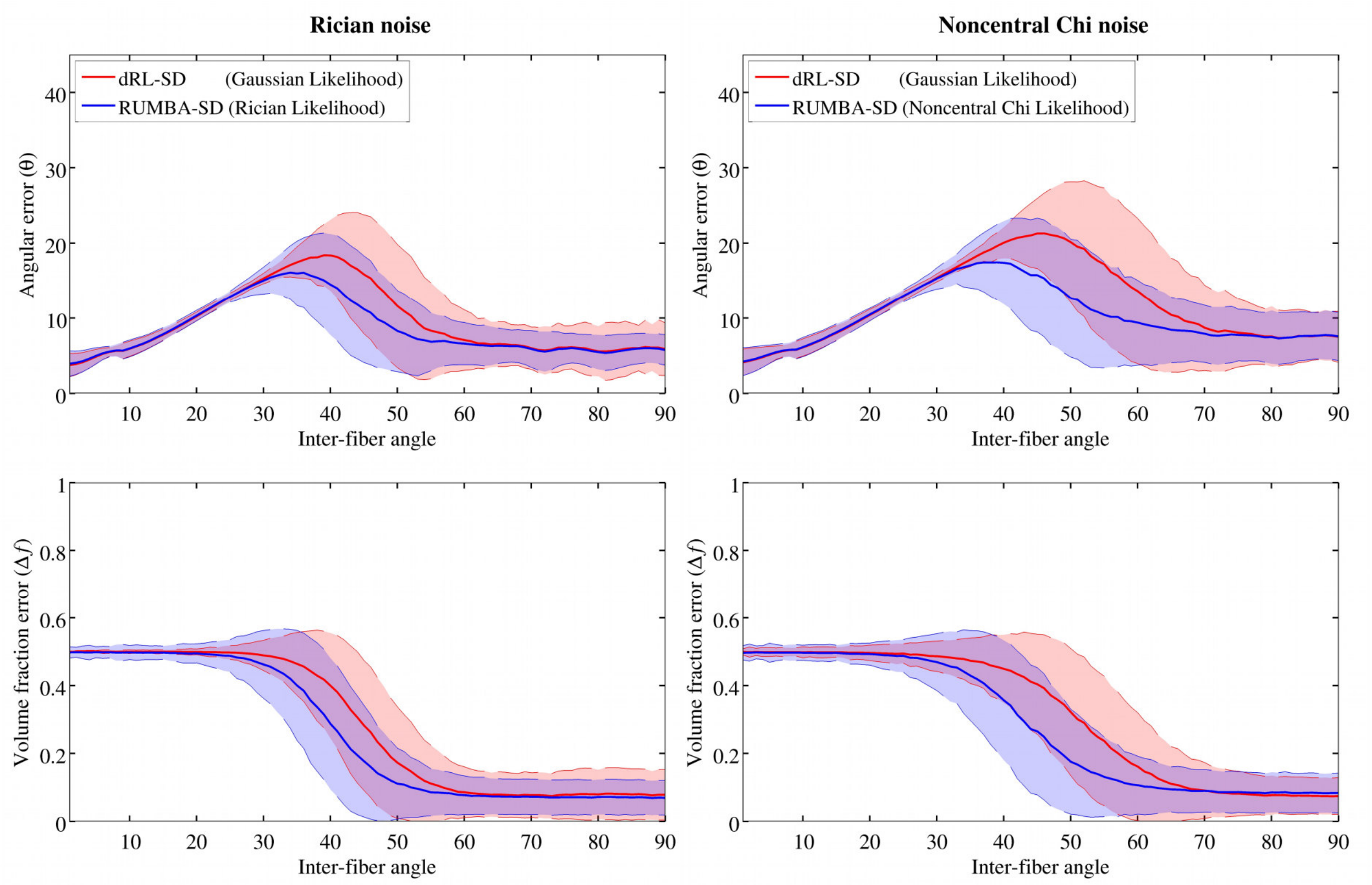

**Fig 1. Reconstruction accuracy for RUMBA-SD and dRL-SD using a dictionary based on original diffusivities**. Reconstruction accuracy of RUMBA-SD (blue color) and dRL-SD (red color) are shown in terms of the angular error ($\theta$) (see Eq.(17)) and the volume fraction error ($\Delta f$) (see Eq. (18)), as a function of the inter-fiber angle in the 90 synthetic phantoms. Continuous lines in each plot represent the mean values for each method. The semi-transparent coloured bands symbolize values within one standard deviation from both sides of the mean. Analyses are based on a dictionary created with the same diffusivities used to generate the data and with a SNR = 15.

A set of patterns can be drawn from these results. First, RUMBA-SD was able to resolve fiber crossings with smaller inter-fiber angles (around 5 degrees and 10 degrees for datasets corrupted with Rician and nc-$\chi$ noise respectively). Second, RUMBA-SD produced volume fraction estimates with a higher precision (lower variance), even in phantoms where the fiber configuration was well-resolved by both methods. Third, although dRL-SD produced a relatively lower proportion of spurious fibers ($n^{+}$), RUMBA-SD produced a lower proportion of undetected fibers ($n^{-}$) leading to a higher success rate (SR). Finally, the performance of both methods was inferior when the dictionary was created using diffusivities estimated from a ‘standard’ DTI fitting in WM regions of parallel fibers. In line with previous findings on dRL-SD [50], optimal results were obtained from the sharper fiber response model.

All these points also hold for results obtained with other SNRs and with a differing number of algorithm iterations. Specifically, when a higher number of iterations was employed (i.e., 400), a lower proportion of $n^{-}$ and a higher proportion of $n^{+}$ was obtained with both methods.

Fig 2 shows the performance of dRL-SD and RUMBA-SD in the 41 phantoms with inter-fiber angle equal to 70 degrees and using different volume fractions. Specifically, average values and standard deviations for the estimated volume fractions of the smaller fiber group are reported for the SNR=15 datasets. Results are based on 200 iterations, using the dictionary created with the sharper fiber response model. Additional results on $n^{+}$ and $n^{-}$ are show in Fig F in S2 File.

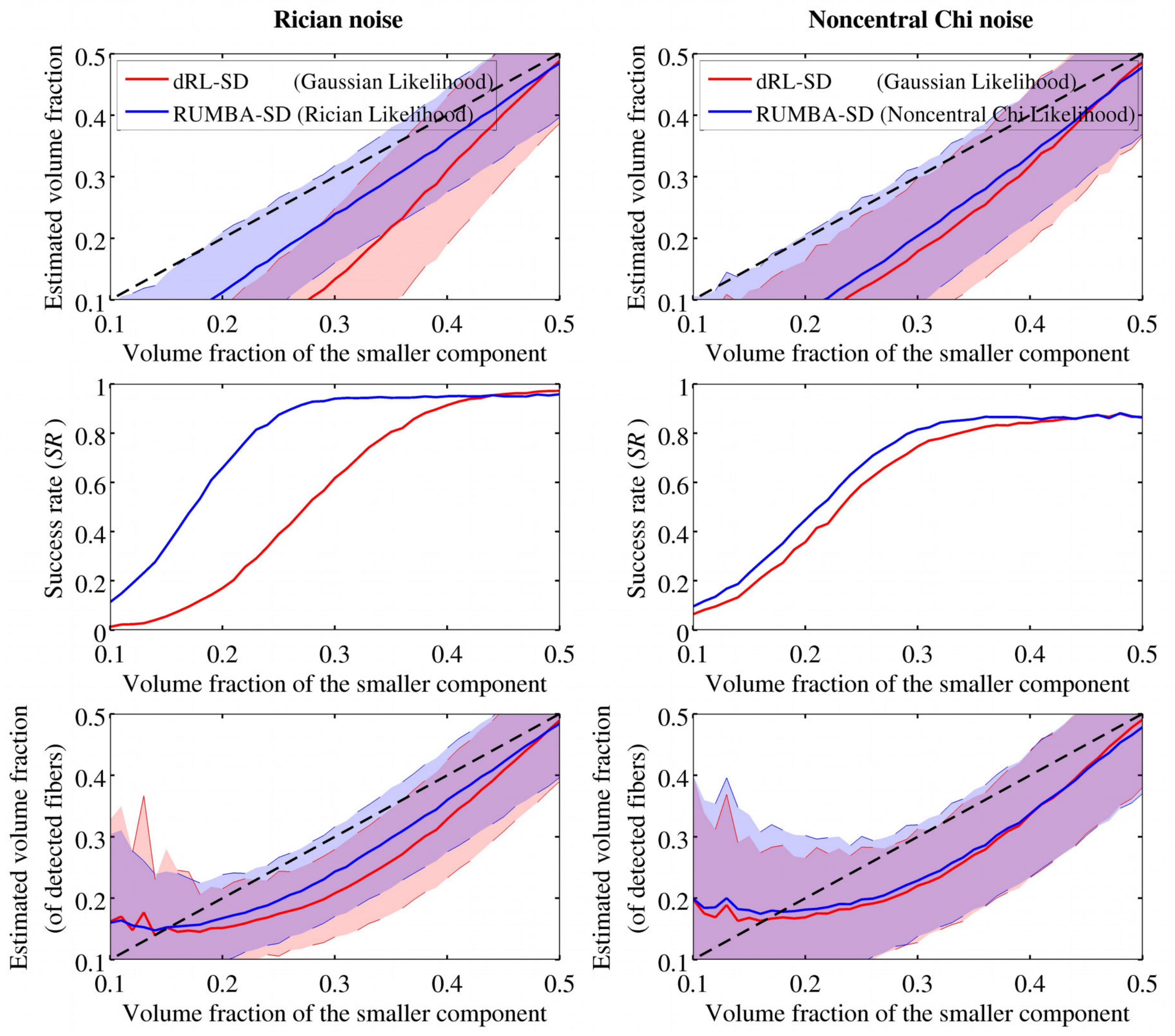

**Fig 2. Reconstruction accuracy of RUMBA-SD and dRL-SD measured in phantoms with different volume fractions.** Reconstruction accuracy of RUMBA-SD (blue color) and dRL-SD (red color) is shown in terms of the volume fraction of the smaller fiber bundle (upper panel) and the success rate (middle panel) in the 41 synthetic phantoms with inter-fiber angle equal to 70 degrees, using different volume fractions. The lower panel shows results similar to those depicted in the upper panel but considering only voxels where the two fiber bundles were detected. The discontinuous diagonal black line in the upper and lower panels represents the ideal result as a reference. The continuous coloured lines in each plot denote the mean values for each method. The semi-transparent coloured bands represent the values within one standard deviation to both sides of the mean. Results refer to the datasets with SNR=15 and dictionary created with the true diffusivities.

The performance of RUMBA-SD was better in terms of the estimated volume fractions and the success rate, especially in the Rician noise case. In order to verify if the lower bias in the volume fractions estimated by RUMBA-SD could be explained only by its higher success rate, the calculation was repeated by considering only those voxels in each phantom where the two fiber populations were identified. After correcting for this factor, a noticeable advantage was still observed for RUMBA-SD (see left lower panel of Fig 2).

## 2 Original versus TV-regularized algorithms

Fig 3 reports angular and volume fraction errors corresponding to the TV spatially-regularized versions of both methods applied to the 90 phantoms characterizing the different inter-fiber angles. Results are based on the same parameters and options used in Fig 1: dictionary created using the sharper fiber response model; noisy datasets with a SNR=15 and reconstructions using 200 iterations. Average values of SR, $n^{+}$ and $n^{-}$ are reported in Fig G in S2 File.

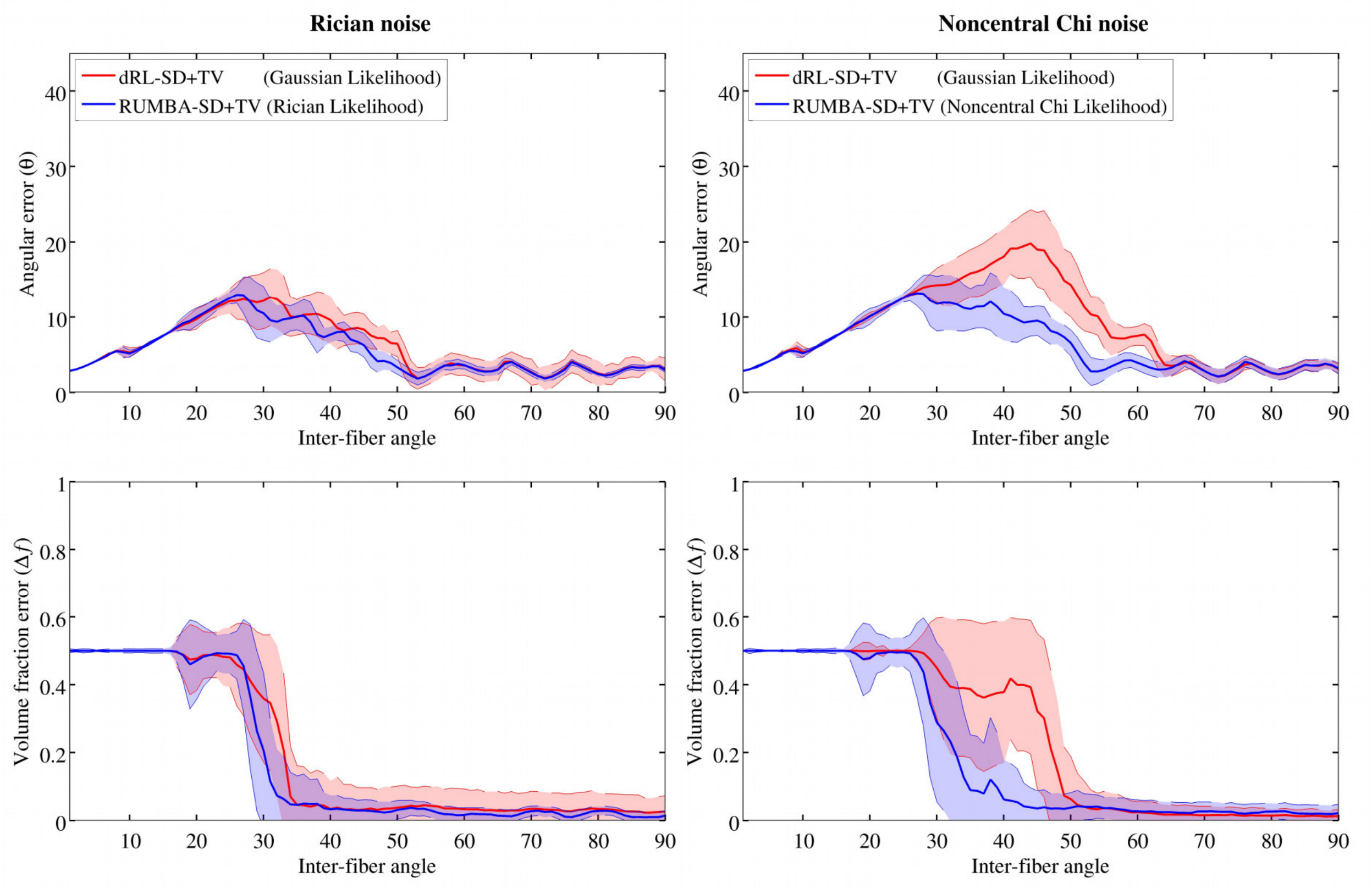

**Fig 3. Reconstruction accuracy levels of RUMBA-SD+TV and dRL-SD+TV**. Reconstruction accuracy of RUMBA-SD+TV (blue color) and dRL-SD+TV (red color) is shown in terms of the angular error ($\theta$) (see Eq. (17)) and the volume fraction error ($\Delta f$) (see Eq. (18)) as a function of the inter-fiber angle in the 90 synthetic phantoms. Continuous lines are the mean values for each method, and semi-transparent coloured bands contain values within one standard deviation on both sides of the mean. This analysis is based on a dictionary created with the same diffusivities used to generate the data with a SNR = 15.

When comparing these results with those from Fig 1 (and Fig D in S2 File), it is clear that TV regularization provides multiple benefits in both algorithms, including a superior ability to detect fiber crossings with smaller inter-fiber angles and a higher success rate. The later is due to a lower proportion of undetected fibers ($n^-$) and of spurious fibers ($n^+$). This pattern is evident in Fig 4, which depicts the peaks extracted from the fiber ODFs estimated from the SMF-based phantom with inter-fiber angle equal to 45 degrees and SNR=15. Peaks are plotted as thin cylinders.

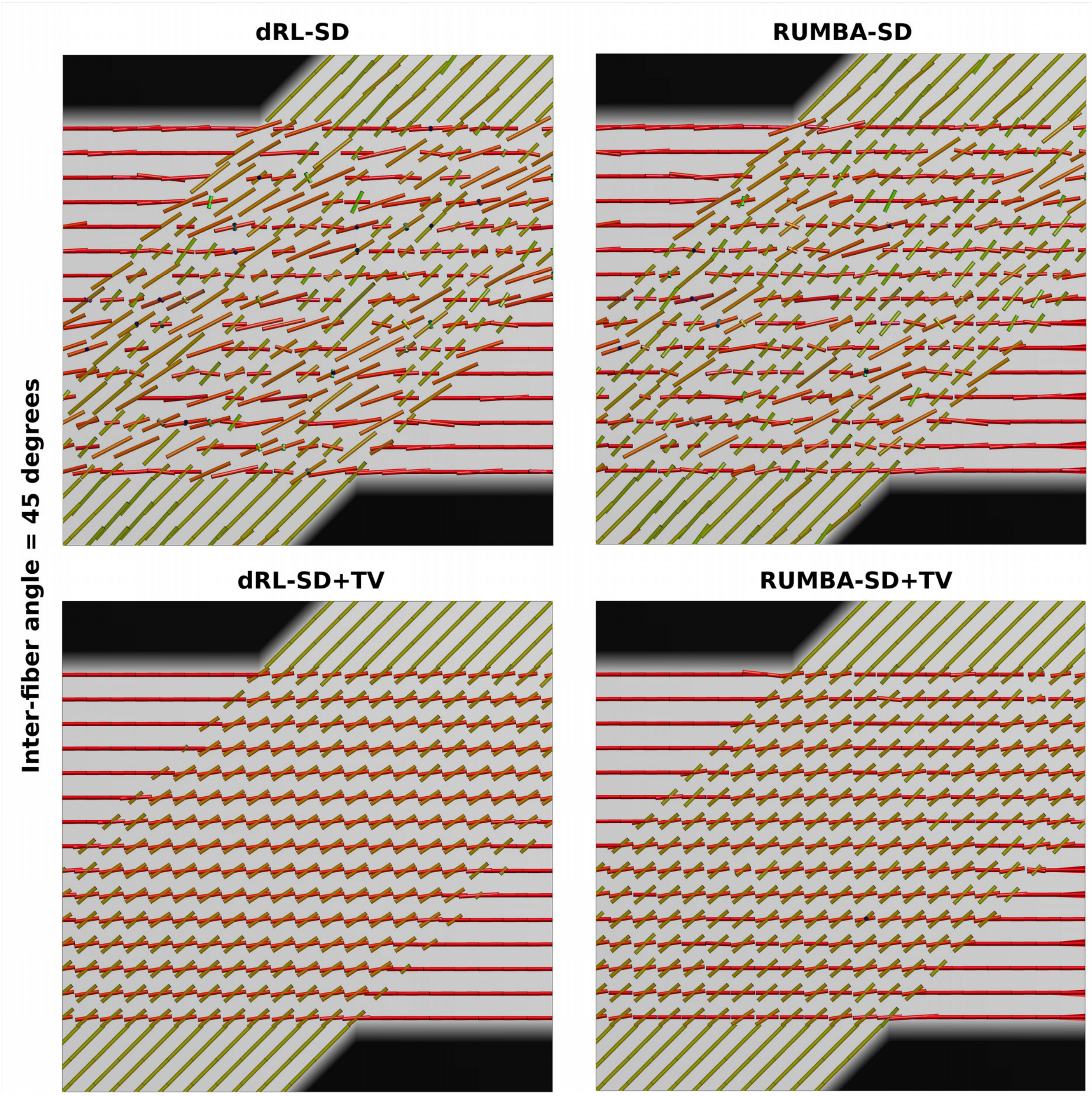

**Fig 4. Main peaks in the 45-degrees phantom data**. Main peaks extracted from the fiber ODFs estimated in the phantom data with inter-fiber angle equal to 45 degrees and Rician noise with a SNR=15 are shown. Results are based on reconstructions using 200 iterations. Peaks are visualized as thin cylinders.

As before, the above patterns were also observed in the analyses obtained with datasets based on other SNRs and with different number of iterations. In all cases RUMBA-SD+TV detected fiber crossings at lower inter-fiber angles. Fig 5 shows an example of this in the phantom with inter-fiber angle equal to 33 degrees corrupted with Rician noise and SNR=15.

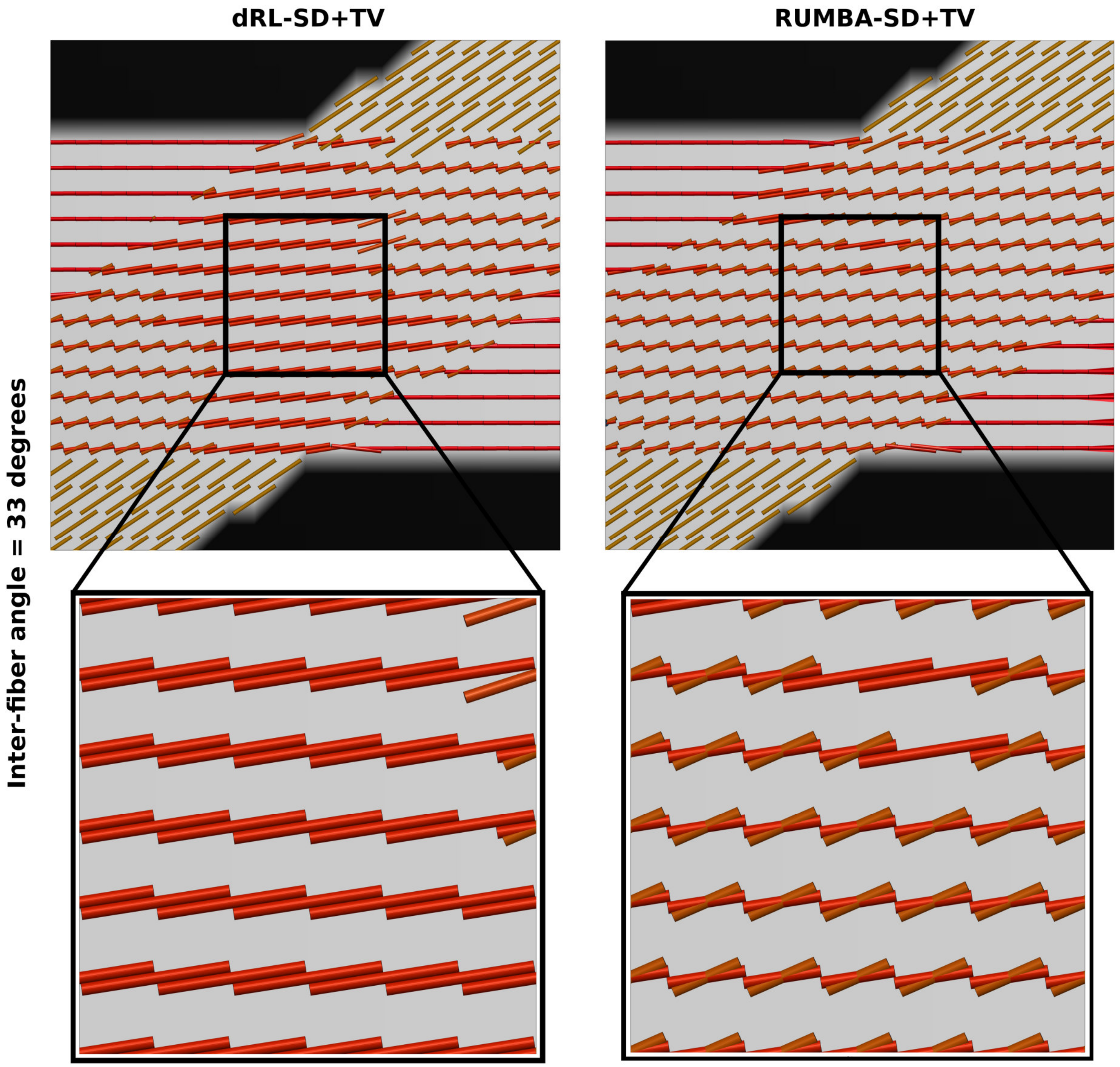

**Fig 5**. **Main peaks in the 33-degrees phantom data**. Main peaks extracted from the fiber ODFs estimated in the phantom data with inter-fiber angle equal to 33 degrees and Rician noise with a SNR=15 are shown. Results are based on reconstructions using 200 iterations. Peaks are visualized as thin cylinders.

# 3 "HARDI Reconstruction Challenge 2013" phantom data

Fig 6 depicts the peaks extracted from the fiber ODFs estimated in a complex region containing various tracts from the SMF-based data generated with a SNR = 20. Results come from reconstructions using 400 iterations and a dictionary with diffusivities equal to $\lambda_1 = 1.6\ 10^{-3}$ $mm^2/s$ and $\lambda_2 = \lambda_3 = 0.3\ 10^{-3}$ $mm^2/s$. Figs H, I, J and K in S2 File show the results from both methods and their regularized versions in the whole slice. Additionally, Fig L in S2 File shows the results corresponding to the reconstructions using 1000 iterations on the same region of interest depicted in Fig 6.

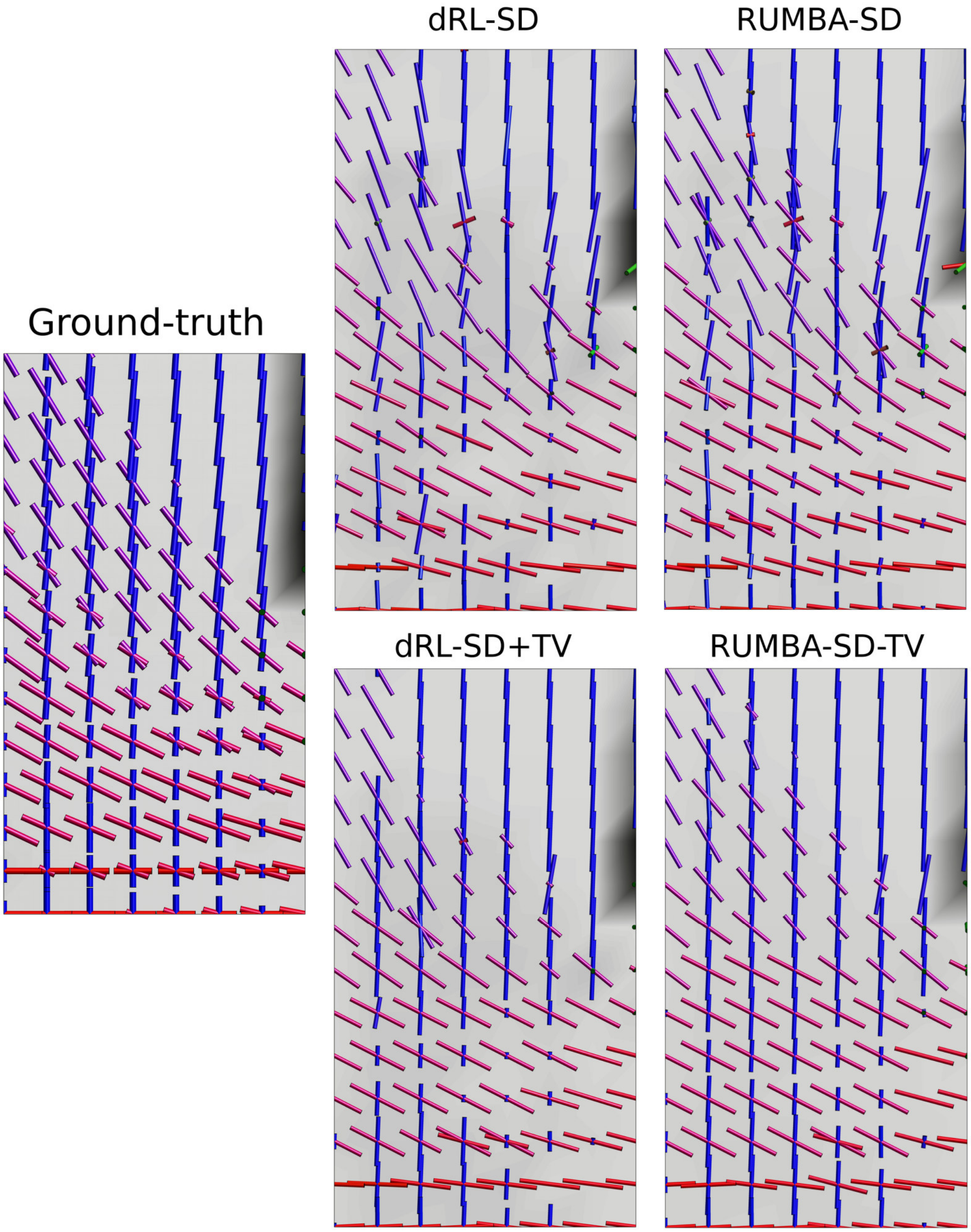

**Fig 6**. **Main peaks from the fiber ODFs estimated in the "HARDI Reconstruction Challenge 2013" phantom**. Visualization of the main peaks extracted from the fiber ODFs reconstructed from the SMF-based data generated with SNR = 20 in a complex region of the "HARDI Reconstruction Challenge 2013" phantom. Results are based on reconstructions using 400 iterations. Peaks are visualized as thin cylinders.

On the one hand, RUMBA-SD was able to resolve some fiber configurations that were not detected by dRL-SD, especially in voxels involving small inter-fiber angles or fiber crossings with a non-dominant tract. On the other hand, the spatially-regularized algorithms have substantially improved the performance of the original methods. Moreover, RUMBA-SD+TV was the method providing better reconstructions. These findings are in line with results from previous sections and remained valid also when different dictionaries, number of iterations and noise levels were employed.

Additional complementary results about the performance of RUMBA-SD in relation to several other reconstruction methods can be found in the website of the 'HARDI Reconstruction Challenge 2013': http://hardi.epfl.ch/static/events/2013_ISBI/workshop.html#results. An earlier version of RUMBA-SD took part in that Challenge, ranking number one in the 'HARDI-like' category (team name: 'Capablanca'). In the discussion section we provide additional information.

# 4 Real brain data

Fiber ODFs were estimated separately for each different SMF- and SoS-based dataset, including the original measured data with the full set of 256 gradient directions and its reduced form including a subset of 64 directions. In all cases, both RUMBA-SD and dRL-SD methods were implemented using the same dictionary, which was created assuming a sharp fiber response model with diffusivities equal to $\lambda_1 = 1.7\ 10^{-3}$ mm$^2$/s and $\lambda_2 = \lambda_3 = 0.3\ 10^{-3}$ mm$^2$/s, and two isotropic terms with diffusivities equal to 0.7 $10^{-3}$ mm$^2$/s and 2.5 $10^{-3}$ mm$^2$/s.

Fig 7 shows the fiber ODFs estimated from the reduced SMF- and SoS-based datasets (i.e., data containing the reduced set of 64 gradient directions) in a coronal ROI on the right brain hemisphere. These results correspond to the reconstructions employing 200 iterations. Visual inspection of Fig 7 reveals that RUMBA-SD has produced sharper fiber ODFs than dRL-SD in both data. It has detected more clearly the fiber crossings. Interestingly, the fiber ODF profiles estimated by dRL-SD from the SoS-based data are smoother than those estimated from the SMF-based data. This behavior is less perceptible in the case of RUMBA-SD, suggesting that it could be more robust to different multichannel combination methods.

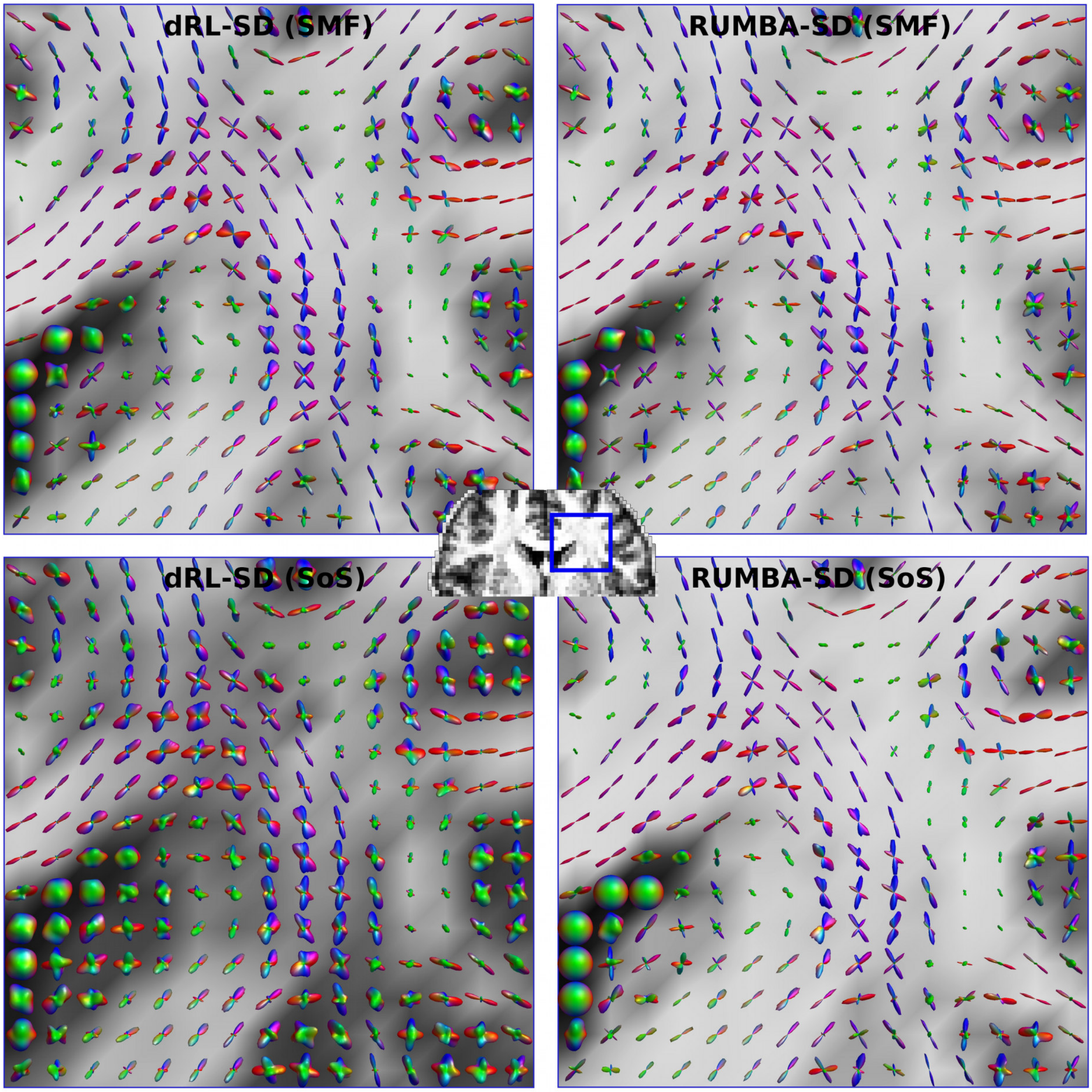

**Fig 7. Fiber ODF profiles estimated from real data**. Visualization of the fiber ODFs estimated in a region of interest on the right brain hemisphere. Results from both SMF- and SoS-based multichannel diffusion datasets (i.e., with Rician and Noncentral Chi noise, respectively) are depicted. The background images are the generalized fractional anisotropy images computed from each reconstruction.

Fig 8 depicts the fiber ODFs estimated with RUMBA-SD and RUMBA-SD+TV in a ROI of both the full and reduced SMF-based datasets. This region contains complex fiber geometries, including the mixture of the anterior limb of internal capsule (alic), the external capsule (ec) and part of the superior longitudinal fasciculus (slf) on the left brain hemisphere. Although in both cases multiple fibers were detected in the area of intersection, RUMBA-SD+TV has provided multi-directional fiber ODFs with a higher number of lobes, which may represent fiber crossings as well as intra-voxel fiber dispersion. The similarity of the reconstructions from the full and reduced datasets suggests that the method is robust with respect to the number of measurements, with the regularized version being the most robust.

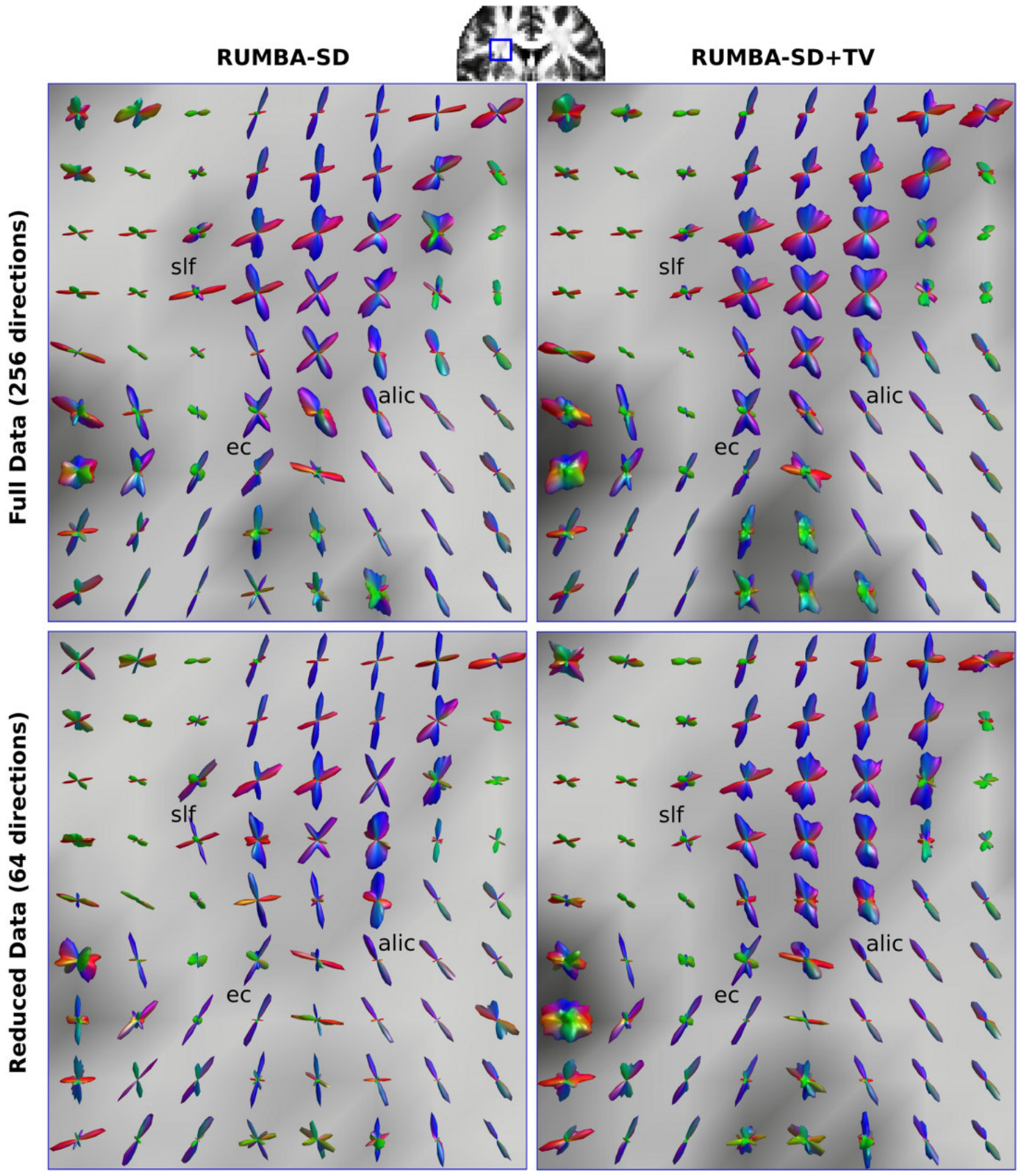

**Fig 8**. **Fiber ODF profiles estimated from real data**. Visualization of the fiber ODFs estimated from RUMBA-SD and RUMBA-SD+TV in a region of interest on the left brain hemisphere. Results are based on estimates employing 300 iterations. The upper and lower panels correspond to results from the full and reduced SMF-based datasets, respectively. The following tracts are highlighted: alic (anterior limb of internal capsule), ec (external capsule), and part of the slf (superior longitudinal fasciculus).

In a subsequent analysis we examined the statistical properties of inter-fiber angles as estimated by all methods. Fig 9 depicts scatter plots of inter-fiber angles estimated by dRL-SD and RUMBA-SD (panel A) and by dRL-SD+TV and RUMBA-SD+TV (panel B). These results are based on reconstructions employing 300 iterations in the 64-direction SMF-based dataset. Only white matter voxels where both methods detected one or two fibers are included in each plot, with inter-fiber angles in single fiber voxels assumed to be zero. Points on the main diagonal line characterize voxels where both methods gave identical inter-fiber angle estimates, whereas points above and below the main diagonal correspond to voxels where the two methods detected two fibers with different inter-fiber angle. The high density of points forming two secondary lines near the main diagonal indicates nearly similar reconstructions by both methods, with the angular differences being similar to the angular resolution of the reconstruction grid (i.e., about 8 degrees). The higher number of points above the main diagonal in both panels, especially for inter-fiber angles lower than 50 degrees, suggests that RUMBA-SD and RUMBA-SD+TV provide higher inter-fiber angles than dRL-SD and dRL-SD+TV, respectively. Finally, points located on the X and Y axes are voxels where one method detected two fibers while the other detected one. The very low density of points on the X axis of panel A for inter-fiber angles lower than 50 degrees (see the blue bracket) indicates that in nearly all voxels where dRL-SD detected two fibers, RUMBA-SD was also able to detect two fibers. In contrast, the high density of points on the Y axis in the same range of inter-fiber angles indicates that in many voxels RUMBA-SD detected two fibers whereas dRL-SD detected a single one. A similar but attenuated effect can be noticed in panel B, suggesting that TV regularization contributes to reduce the differences between both methods.

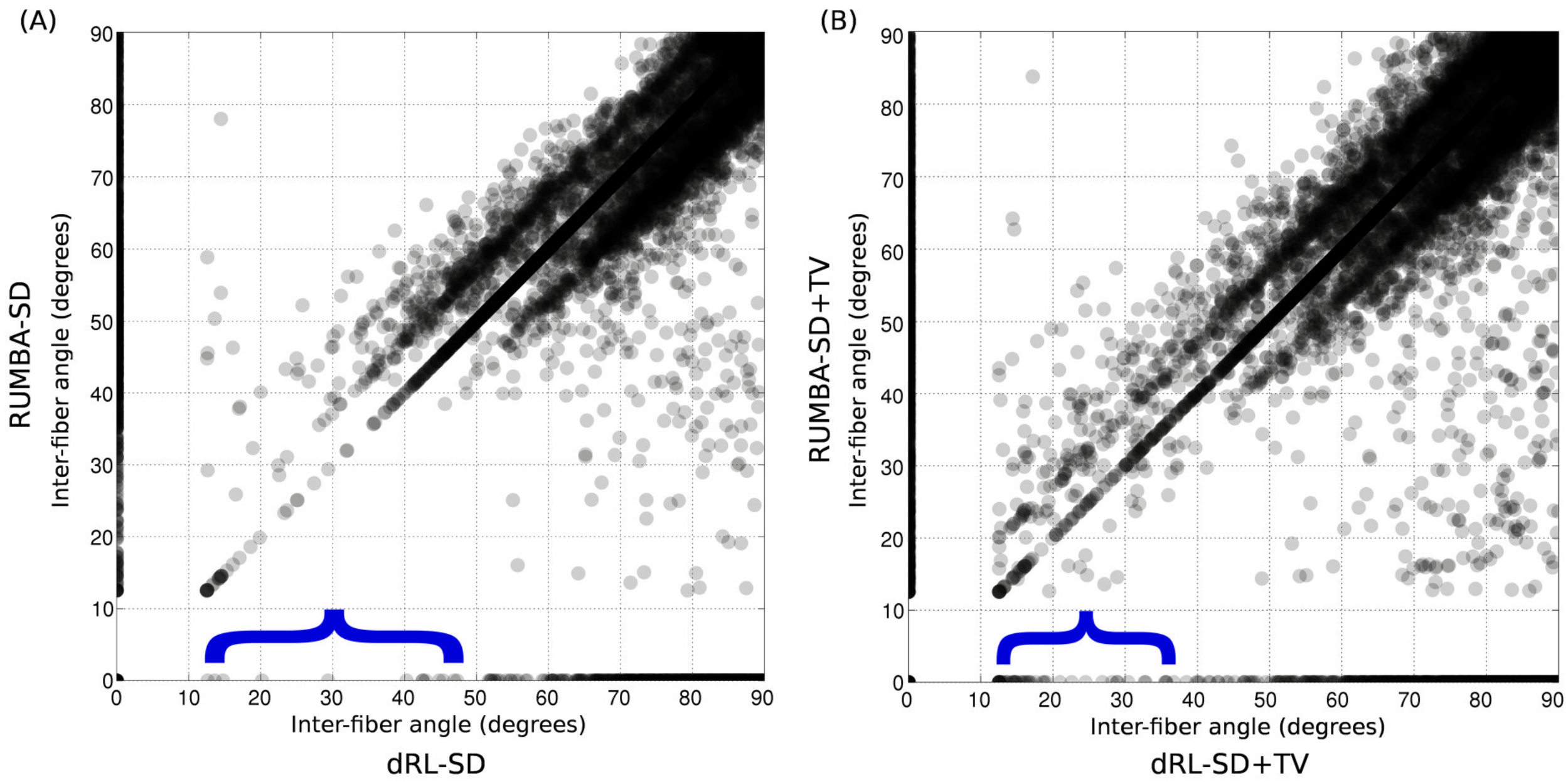

**Fig 9**. **Scatter plots of inter-fiber angles estimated in real data**. Scatter plots of the inter-fiber angles estimated by dRL-SD and RUMBA-SD (panel A) and by dRL-SD+TV and RUMBA-SD+TV (panel B)

in the same voxels. Results are based on reconstructions in the 64-direction SMF dataset. Only voxels in white matter where both methods detected one or two fibers are included. The inter-fiber angle in voxels with a single fiber was assumed to be equal to zero. Points on the main diagonal line are those voxels where the inter-fiber angle estimated from both methods was identical, whereas points above and below the main diagonal correspond to voxels where the two methods detected two fibers but with different inter-fiber angle. Points located on the X and Y axes are voxels where one method detected two fibers whereas the other detected a single fiber.

# Discussion and Conclusions

In this study we propose a new model-based spherical deconvolution method, RUMBA-SD. In contrast to previous methods, usually based on Gaussian noise with zero mean, RUMBA-SD considers Rician and noncentral Chi noise models, which are more adequate for characterizing the non-linear bias introduced in the diffusion images measured in current 1.5T and 3T multichannel MRI scanners. Although recent progress has been made in new SD methods adapted to corrupted Rician data, e.g., see [41, 76] to the best of our knowledge, our study provides the first SD extension to noncentral Chi noise. Furthermore, RUMBA-SD offers a very general estimation framework applicable to different datasets, with its flexibility emanating from two features: i) the explicit dependence between the likelihood model and the number of coils in the scanner and ii) the specific methodology employed to combine multichannel signals. In addition, the voxel-wise estimation of the noise variance adequately deals with potential deviations in the noise distribution, due, for instance, to accelerated MRI techniques or to preprocessing effects. We hope that the proposed technique will help extend SD methods to a wide range of datasets taken from different scanners and using different protocols.

This study adds to previous diffusion MRI studies trying to overcome the signal-dependent bias introduced by Rician and noncentral Chi noise. Apart from the robust DTI estimation methods in [77] and the earlier DTI study conducted by [78], the noise filtering techniques recently described in [79, 80] are specially relevant. These techniques can be applied in the pre-processing steps prior to HARDI data estimation.

The performance of RUMBA-SD has been evaluated exhaustively against the state-of-the-art dRL-SD technique. For that, we have used 132 different 3D synthetic phantoms, including 90 phantoms simulating fiber crossings with different inter-fiber angle, 41 phantoms simulating fiber crossings with different volume fraction, and the complex phantom designed for the "HARDI Reconstruction Challenge 2013" Workshop organized within the IEEE International Symposium on Biomedical Imaging. The comparison of these two methods has allowed us to weight the impact on the results of the Rician and noncentral Chi likelihood models included in RUMBA-SD, in relation to the Gaussian model assumed in dRL-SD. Since both approaches were implemented using the same dictionary of basis signals and similar reconstruction methods based on Richardson-Lucy algorithms adapted to Gaussian, Rician and noncentral Chi noise models, results should be considered comparable.

Taken together, findings from all synthetic datasets demonstrate the benefits of an adequate modelling of the noise distribution in the context of spherical deconvolution, and of the inclusion of TV regularization. Interestingly, RUMBA-SD resolved fiber crossings with smaller inter-fiber angles and smaller non-dominant fibers. Likewise, RUMBA-SD produced volume fraction estimates with higher accuracies and precision and produced, as well, a lower proportion of undetected fibers, resulting in a higher success rate (see Figs 1 and 2 and supplementary figures in S2 File). On the other hand, the TV spatially-regularized versions of both dRL-SD and RUMBA-SD have substantially improved the performance of the original methods in all studied metrics (see Figs 3-6 and supplementary figures in S2 File).

As previously mentioned, an earlier version of RUMBA-SD took part in the HARDI Reconstruction Challenge 2013, ranking number one in the 'HARDI-like' category. Notably, this position was shared with a reconstruction based on the CSD method included in *Dipy* software [81] (http://nipy.org/dipy/), which had applied a Rician denoising algorithm [82] to the raw diffusion MRI data prior to the actual CSD reconstruction. The superior performance of these two approaches strengthens the importance of taking into account the non-Gaussian nature of the noise. Moreover, it opens new questions on the optimal strategy to be followed. Should we divide the deconvolution process into to two disjoint steps (first denoising and then estimation) or it is more adequate to follow the unified approach proposed here?. The main advantage of the former is that it may benefit from including state-of-art denoising algorithms like the adaptative non-local means denoising method proposed in [82]. Conversely, the main advantage of the unified approach is that it provides a precise model to distinguish real signals from noise throughout all the 4D diffusion MRI data. Many of the advanced denoising algorithms that are currently applied in isolation were developed to filter volumetric (3D) data. Since each 3D image is processed individually, their mutual dependence in terms of orientation is ignored. In contrast, the unified approach described here provides a more general estimation framework that may be extended to include advanced similarity measures like those employed in [82], merging the benefits of both strategies. A new manuscript on the 'Challenge' phantom (currently under preparation) will provide additional information on the performance of RUMBA-SD in relation to several reconstruction methods and in terms of connectivity metrics derived from fiber tracking analyses.

When applied to human brain data, RUMBA-SD has also achieved the best results, with its reconstructions showing the highest ability to detect fiber crossings (see Figs 7-9). And although any conclusion derived from real data is hampered by the unknown anatomy at the voxel level, all previous results on synthetic data seem to support the validity of RUMBA-SD for real data.

Our findings can also be contrasted with those reported in [55]. In that study, the authors show that the SoS approach produces a signal-dependent bias that reduces the signal dynamic range and may subsequently lead to decreased precision and accuracy in fiber orientation estimates. Our study, however, suggests that the noncentral Chi noise in SoS-based data is not a major concern for the SD methods considered. Thus, heavier squashing of fiber ODFs when SoS reconstruction is used [55] is not that prominent with the SD if compared to diffusion ODF estimation methods [11]. This result may have different explanations for each technique. The robustness of dRL-SD may be explained, in part, by its lower over-all sensitivity to selection of the response function [50], which make it robust to the use of dictionaries estimated from either biased or unbiased signals. This behaviour may be additionally boosted by the inclusion of the damping factor in the RL algorithm. In contrast, the robustness of RUMBA-SD can be explained by the use of proper likelihood models that explicitly consider the bias as a function of the noise corrupting the data.

To finish, some limitations and future extensions of the study should be acknowledged. First, we have not evaluated the proposed method in synthetic data simulating partial Fourier *k*-space acquisitions and with parallel imaging using various acceleration factors (i.e., R>1), yet it may be interesting to consider it. Second, the RUMBA-SD estimation framework is based on a discrete approximation of the fiber ODF, which may be potentially extended to continuous functions on the sphere, like the spherical harmonics and the wavelets. Third, the TV regularization implemented in this study is based on a channel-by-channel first order scheme. New studies may be designed to compare different regularization techniques such as higher order TV, vectorial TV and the fiber continuity approach introduced in [56], to mention only a few examples. Fourth, different strategies for creating the signal dictionary could be explored, like using mixtures of intra-compartment models to capture different diffusion profiles, or applying more appropriate models to fit multi-shell data [31, 83]. Fifth, the recursive calibration of the single-fiber response function proposed by [84] may be another possible add-on. Finally, it is worth mentioning that the inversion algorithm behind RUMBA-SD is not limited to fiber ODF reconstructions, but it can be also applied to solve other linear mixture models from diffusion MRI data. It was recently showed that some microstructure imaging methods such as ActiveAx and NODDI can be reformulated as convenient linear systems, however, the deconvolution methods proposed for them assume Gaussian noise and are performed on a voxel-by-voxel basis [85]. Here the iterative scheme proposed in RUMBA-SD could be used to address both limitations, potentially leading to improved reconstructions in microstructure imaging.

# Acknowledgments

We would like to thank Dr Karla Miller for assisting us with data collection. The presented study is a tribute to composers who popularized Rumba music in the 1940s and 50s, including Dámaso Pérez Prado, Mongo Santamaría, Xavier Cugat, and Chano Pozo.

# Supporting Information

**S1 File. Supplementary appendices**

**S2 File. Supplementary figures**

# S1 File. Supplementary appendices

### A) Relationship with the RL-SD method

In the limit of very high signal to noise ratio (i.e., $\mathbf{S} \circ \mathbf{H}\mathbf{f}^k / \sigma^2 \gg n$) the modified Bessel functions ratio in Eq. (8) tends to the unity (see Fig M in **S2 File**). In that limit, Eq. (8) becomes:

$$\mathbf{f}^{k+1} = \mathbf{f}^k \circ \frac{\mathbf{H}^T \mathbf{S}}{\mathbf{H}^T \mathbf{H} \mathbf{f}^k}, \qquad \textbf{(A-1)}$$

which is just the undamped RL-SD method originally proposed in [42] under the assumption of zero-mean Gaussian noise .

### B) Noncentral Chi noise model for SoS data - effective or standard values?

As previously discussed, for scanners with a high number of coils where the effect of noise correlation cannot be easily decoupled, a best approximation for the noise model in SoS-based images is obtained by using effective $n_{eff}$ and $\sigma_{eff}^2$ values [63]. This section is devoted to provide an initial insight on the implication of using a noncentral Chi noise model in our estimation with standard parameters, instead of the effective ones.

If considering the limit $n \gg 1$, the ratio provided by Eq. (C-1) in Appendix C can be approximated as:

$$\frac{I_n(x)}{I_{n-1}(x)} \approx \frac{x}{n + \sqrt{x^2 + n^2}}, \qquad \textbf{(B-1)}$$

where we have used the identity relating the expansion of a square root in terms of continued fraction. This expression can be regarded as a lower bound for the true ratio, which is more accurate insofar as $n$ increases.

Notably, based on this result we obtain

$$\frac{I_n\left(\mathbf{S}\circ\mathbf{Hf}/\sigma^2\right)}{I_{n-1}\left(\mathbf{S}\circ\mathbf{Hf}/\sigma^2\right)} \approx \frac{\mathbf{S}\circ\mathbf{Hf}}{n\sigma^2+\sqrt{\left(\mathbf{S}\circ\mathbf{Hf}\right)^2+\left(n\sigma^2\right)^2}}. \quad \textbf{(B-2)}$$

Similarly,

$$\frac{I_{neff}\left(\mathbf{S}\circ\mathbf{Hf}/\sigma_{eff}^2\right)}{I_{neff-1}\left(\mathbf{S}\circ\mathbf{Hf}/\sigma_{eff}^2\right)} \approx \frac{\mathbf{S}\circ\mathbf{Hf}}{n_{eff}\sigma_{eff}^2+\sqrt{\left(\mathbf{S}\circ\mathbf{Hf}\right)^2+\left(n_{eff}\sigma_{eff}^2\right)^2}}. \quad \textbf{(B-3)}$$

The relevant feature of these relationships is that they do not depend on the individual parameters of interest but just on their products $n\sigma^2$ and $n_{eff}\sigma_{eff}^2$. This implies that although in general the functions $I_n\left(\mathbf{S}\circ\mathbf{Hf}/\sigma^2\right)$ and $I_{neff}\left(\mathbf{S}\circ\mathbf{Hf}/\sigma_{neff}^2\right)$ are different, if $n_{eff}\sigma_{eff}^2 = n\sigma^2$ then their ratios (which are the terms used in the computation) satisfy:

$$\frac{I_n\left(\mathbf{S}\circ\mathbf{Hf}/\sigma^2\right)}{I_{n-1}\left(\mathbf{S}\circ\mathbf{Hf}/\sigma^2\right)} \approx \frac{I_{neff}\left(\mathbf{S}\circ\mathbf{Hf}/\sigma_{eff}^2\right)}{I_{neff-1}\left(\mathbf{S}\circ\mathbf{Hf}/\sigma_{eff}^2\right)}. \quad \textbf{(B-4)}$$

The accuracy of this approximation is determined by the accuracy of Eq. (B-1) and the assumption $n_{eff}\sigma_{eff}^2 = n\sigma^2$. Interestingly, in [54] was reported that for a system with 32 receiver channels of non-accelerated SoS-based data, the mean effective number of channels was $n_{eff}$ =12. Theoretical calculations in [63] predict similar $n_{eff}$ values for 32 and 16 coil systems with correlation coefficients between coil of $\rho \approx 0.3$ and $\rho \approx 0.2$, respectively. In a complementary analysis we have verified that Eq. (B-1) provides a 'reasonable' approximation for that effective number of channels. Moreover, in [73] was showed that for SoS reconstructions without using fast pMRI techniques the product $n_{eff}\sigma_{eff}^2$ is constant across the image and equal to $n\sigma^2$. These results, taken together, indicate that for *non-accelerated* SoS-based data acquired in multichannel scanners with a high number of coils and moderate correlation coefficients between coils the SD estimation process may be approximately performed using the standard parameters. That is, by working in terms of the real parameters $n$ and $\sigma^2$, we can avoid the complex estimation of the spatial-dependent effective parameters.

## C) A note on the evaluation of the term $I_n(x)/I_{n-1}(x)$

The proposed SD algorithm involves the evaluation of the ratio of modified Bessel functions of first kind. Such evaluation is best computed by considering the ratio as a new composite function, and not by means of the simple evaluation of the ratio of the individual functions. The main reason for this is related to the divergence towards infinity of the individual functions. For instance, in *Matlab* software, numerical values for $I_n(x)$ are only available for $x \le 700$; for higher values an infinity value is returned, and thus the ratio cannot be computed.

Interestingly, this ratio can be expressed in terms of Perron continued fraction [70]:

$$\frac{I_n(x)}{I_{n-1}(x)} = \cfrac{x}{2n+x-\cfrac{2x(n+1/2)}{2n+1+2x-\cfrac{2x(n+3/2)}{2n+2+2x-\cfrac{2x(n+5/2)}{2n+3+2x-\ddots}}}} \qquad \textbf{(C-1)}$$

A study about the convergence of this expansion [70] revealed that it converges faster than other analogous representation based on Gauss continued fraction. For the purpose of this application, the summation in Eq. (C-1) is performed up to the final term $\frac{2x(n+5/2)}{2n+3+2x}$. Fig M in **S2 File** shows the accuracy of this approximation for different values of $n$ and for a wide range of values of $x$.

**D) dRL-SD+TV**

The dRL-SD method is based on the following iterative scheme [37]:

$$\mathbf{f}^{k+1} = \mathbf{f}^{k} \circ \mathbf{L}^{k}, \tag{D-1}$$

where

$$\mathbf{L}^{k} = \left[ 1 + \underbrace{\left( 1 - \mu \left( 1 - \frac{\left(\mathbf{f}^{k}\right)^{\nu}}{\left(\mathbf{f}^{k}\right)^{\nu} + \eta^{\nu}} \right) \right)}_{\mathbf{u}} \circ \left( \frac{\mathbf{H}^{T}\mathbf{S} - \mathbf{H}^{T}\mathbf{H}\mathbf{f}^{k}}{\mathbf{H}^{T}\mathbf{H}\mathbf{f}^{k}} \right) \right] \tag{D-2}$$

and $\mu$ is a parameter that depends on the standard deviation of the vector of measurements, $\mu = \max\left(0, 1 - 4std\left(\mathbf{S}\right)\right)$. Notice that when the term $\mathbf{u}$ highlighted in (D-2) is equal to 1, the resulting damped version becomes equal to the original one described in (C-1).

The TV spatially-regularized extension to this method proposed in this work is based on the following modification:

$$\mathbf{f}^{k+1} = \mathbf{f}^{k} \circ \mathbf{L}^{k} \circ \mathbf{R}^{k}, \tag{D-3}$$

where $\mathbf{R}^{k}$ is computed via Eq. (12) and the regularization parameter $\alpha_{TV}$ is computed following an approach similar to that used in RUMBA-SD+TV. Likewise, the noise variance is obtained by minimizing the negative Gaussian log-likelihood with respect to $\sigma^2$, which yields an iterative scheme analogous to Eq. (9).

# S2 File: Supplementary figures

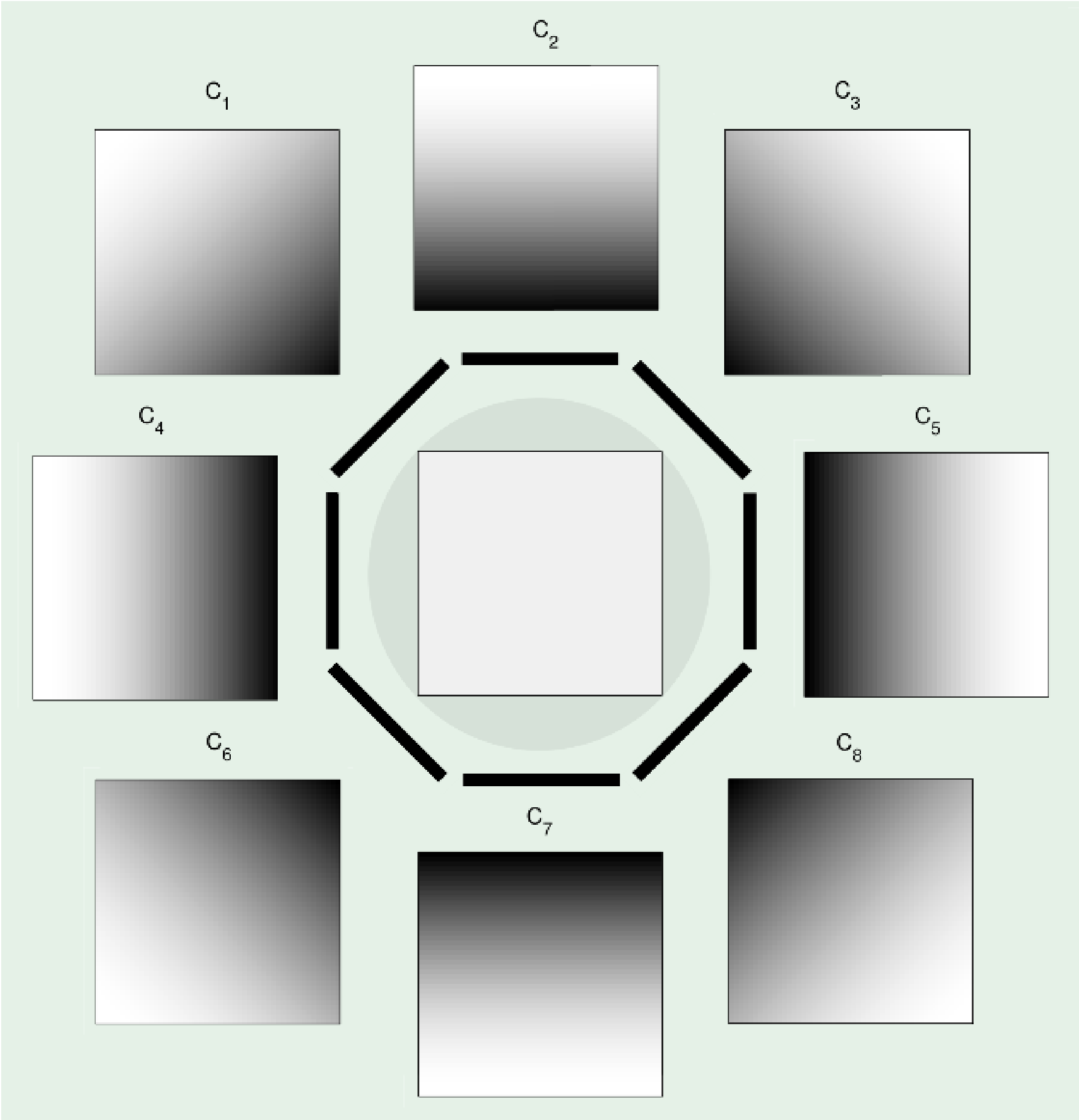

**Fig A**. Sensitivity maps simulating an eight-coil system. White colors denote higher values. The constant image in the center is the sum-of-squares of the individual sensitivity maps.

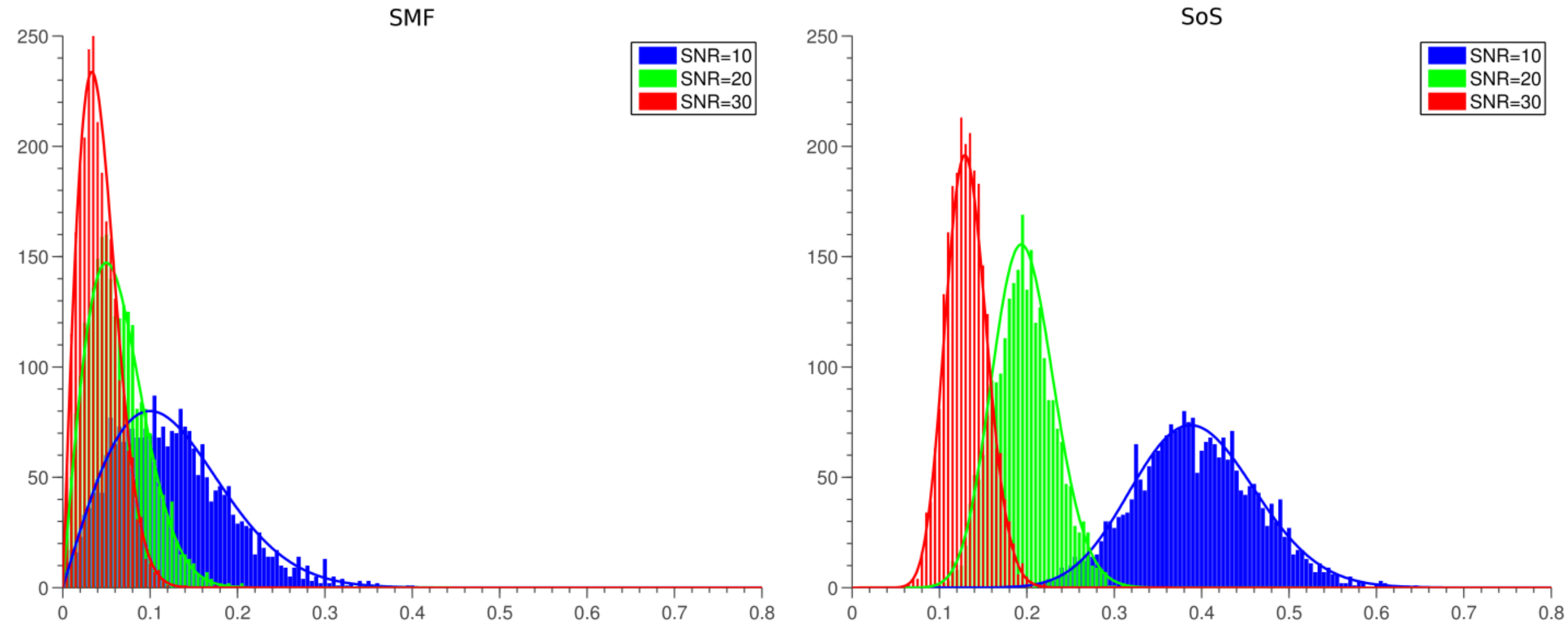

**Fig B**. Noise distribution profiles resulting from background regions outside the "HARDI Reconstruction Challenge 2013" phantom for the two reconstruction methods and SNR =10, 20 and 30. Rician and noncentral Chi distributions are obtained in SMF- and SoS-based data, respectively. The mean value of the distributions increases as long as the SNR decreases. This bias is higher in the SoS data.

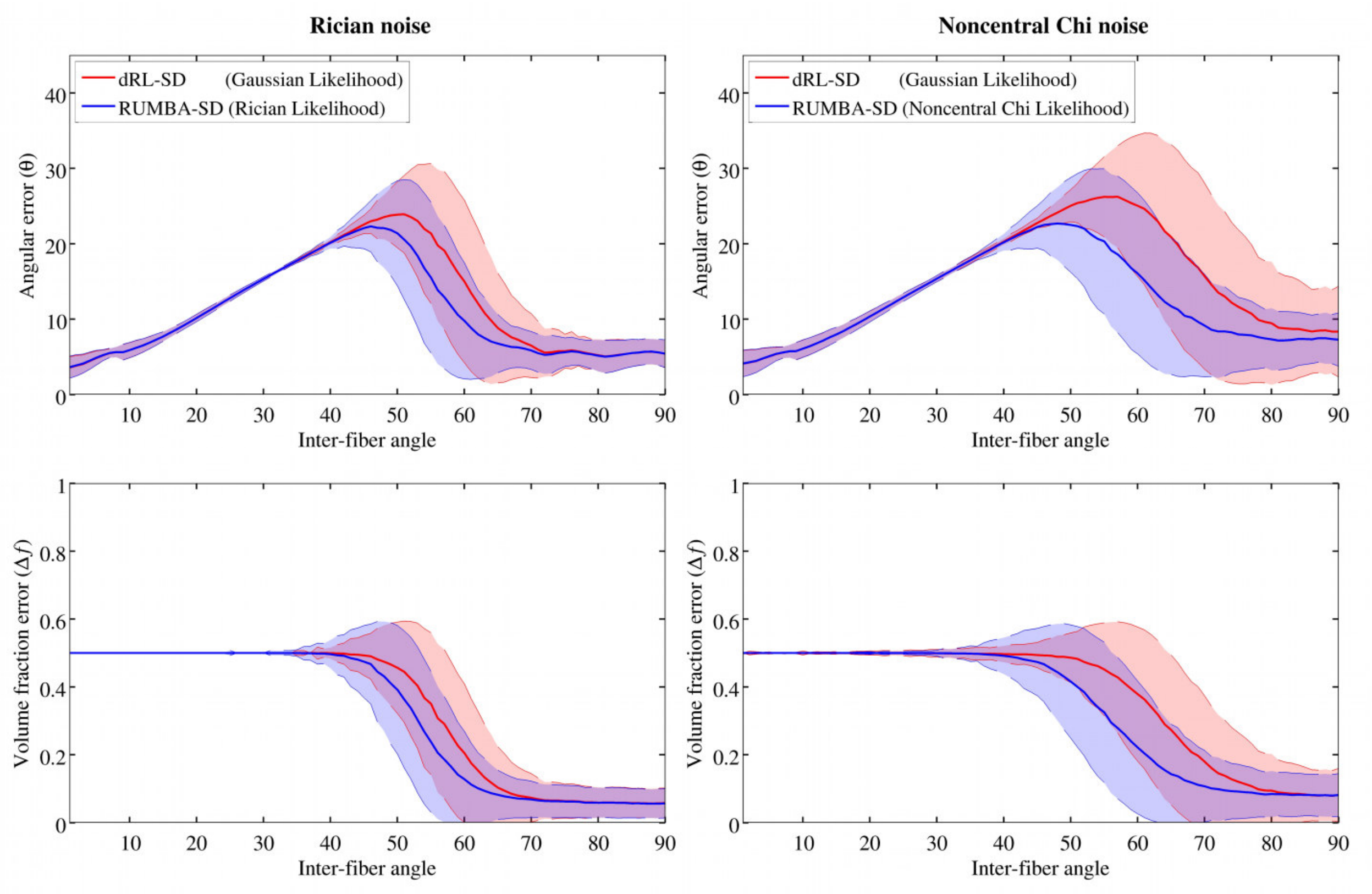

**Fig C.** Reconstruction accuracy levels for RUMBA-SD and dRL-SD using a dictionary based on estimated diffusivities. Reconstruction accuracy of RUMBA-SD (blue color) and dRL-SD (red color) is shown in terms of the angular error ( $\theta$ ) (see Eq. (17)) and the volume fraction error ( $\Delta f$ ) (see Eq.(18)) as a function of the inter-fiber angle in the 90 synthetic phantoms. Continuous lines are mean values and semi-transparent coloured bands contain values within one standard deviation from the mean. Results are based on a dictionary created with empirical diffusivities from a diffusion tensor model applied on the data with a SNR = 15.

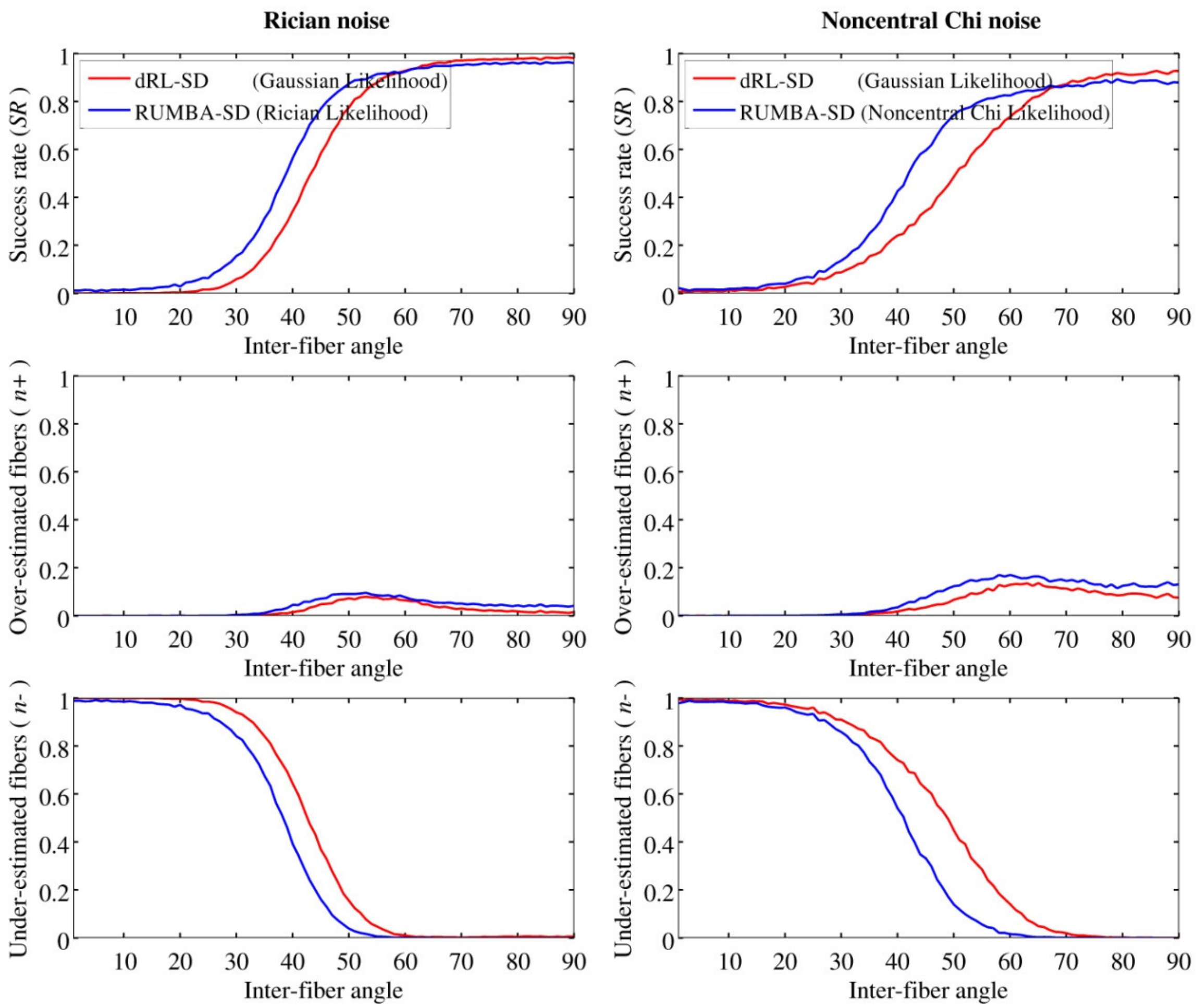

**Fig D**. Quantification of the reconstruction accuracy of RUMBA-SD (blue color) and dRL-SD (red color) in terms of the success rate (SR) and the mean number of over-estimated ( $n^+$ ) and under-estimated ( $n^-$ ) fiber populations, as a function of the inter-fiber angle in the 90 synthetic phantoms. The continuous lines in each plot represent the mean values for each method. This analysis refers to results using a dictionary created with the same diffusivities utilized to generate the data and a level of noise with a SNR = 15.

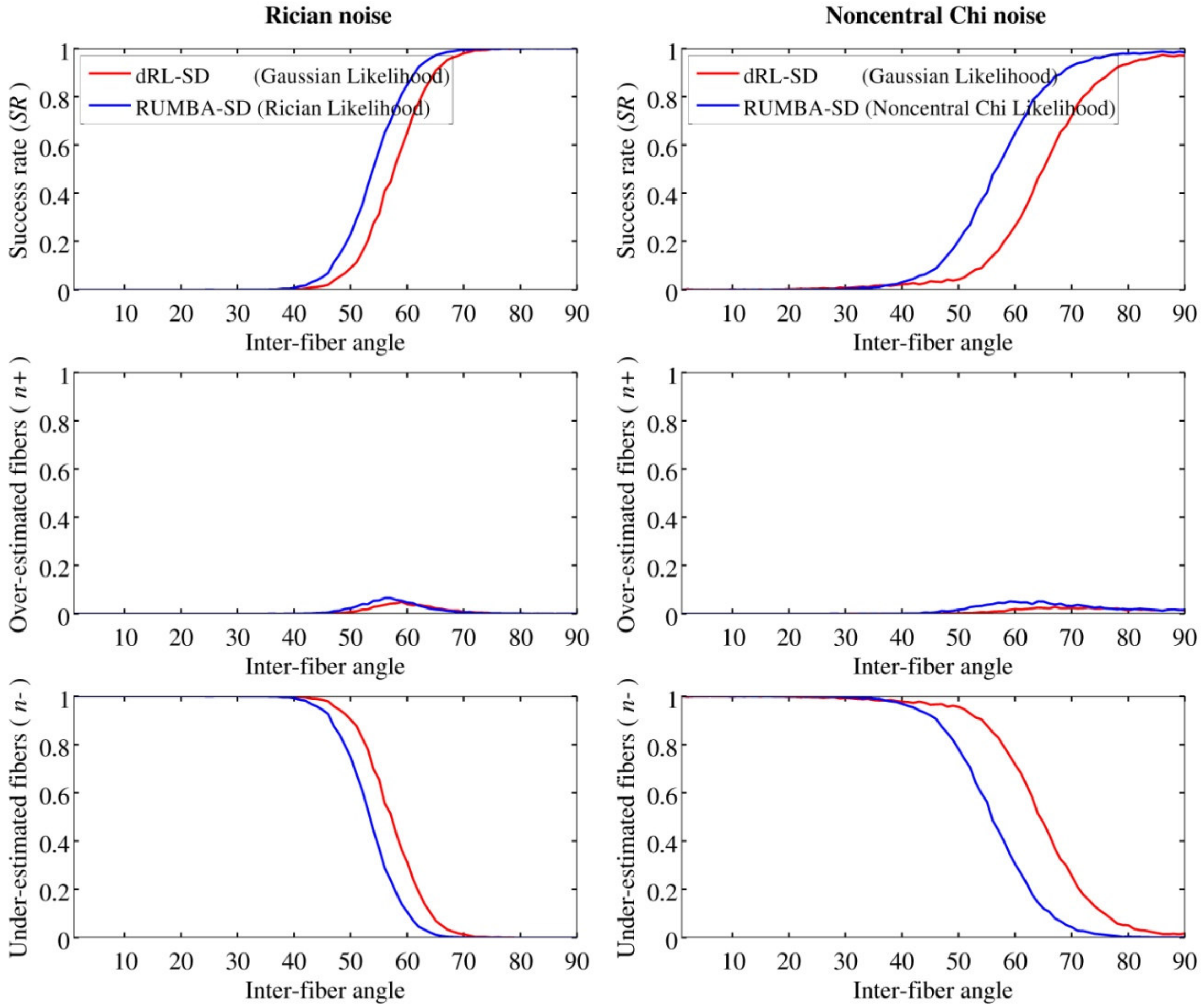

**Fig E**. Quantification of the reconstruction accuracy of RUMBA-SD (blue color) and dRL-SD (red color) in terms of the success rate (SR) and the mean number of over-estimated ( $n^+$ ) and under-estimated ( $n^-$ ) fiber populations, as a function of the inter-fiber angle in the 90 synthetic phantoms. The continuous lines in each plot represent the mean values for each method. This analysis refers to results using a dictionary created with empirical diffusivities estimated by means of the diffusion tensor model from the noisy data with a SNR = 15.

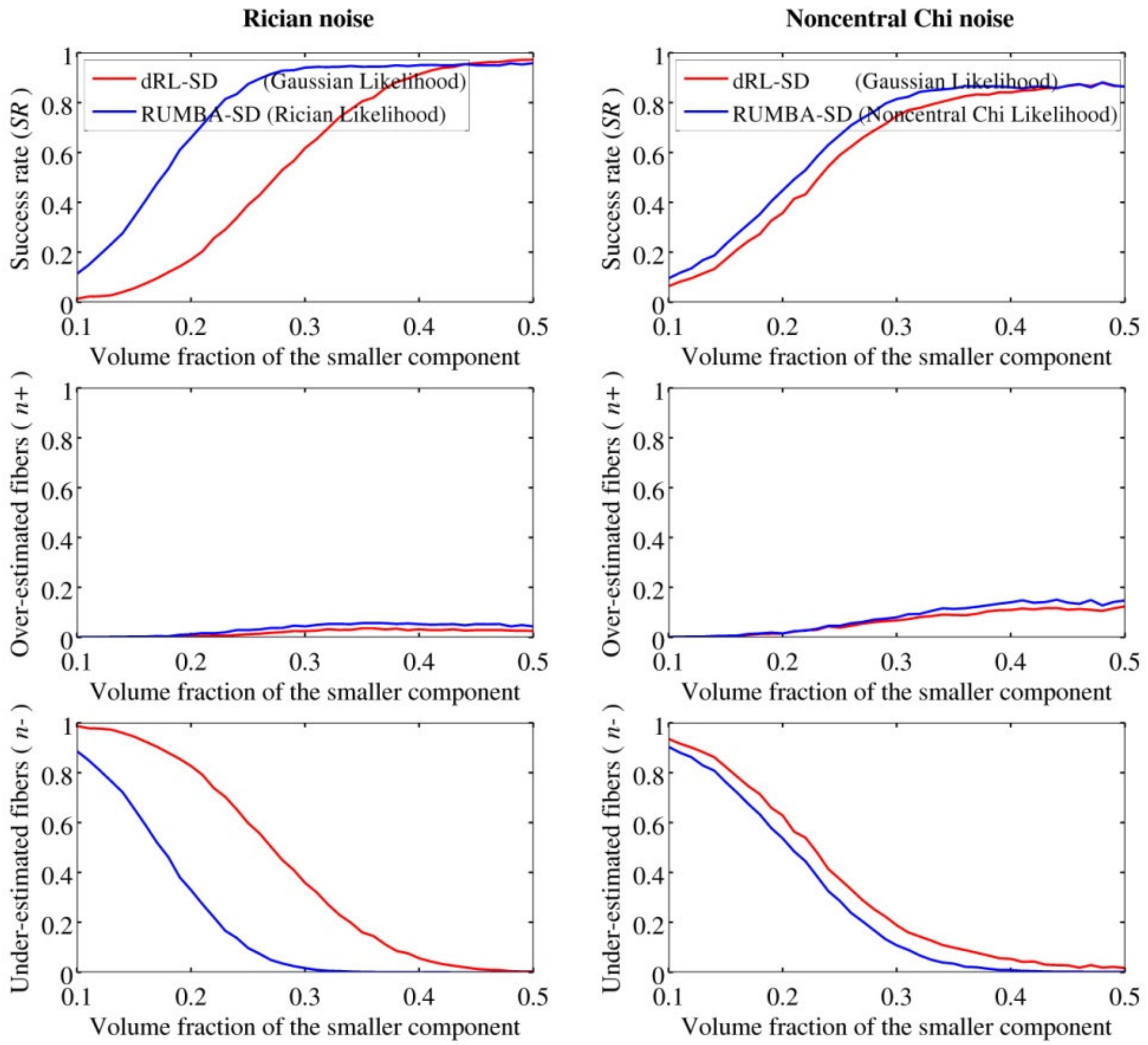

**Fig F**. Quantification of the reconstruction accuracy of RUMBA-SD (blue color) and dRL-SD (red color) in terms of the success rate (SR) and the mean number of over-estimated ($n^+$) and under-estimated ($n^-$) fiber populations, as a function of the volume fraction of the smaller fiber bundle in the 41 synthetic phantoms with inter-fiber angle equal to 70 degrees and different volume fractions. Results refer to the datasets with SNR=15 and dictionary created with the true diffusivities.

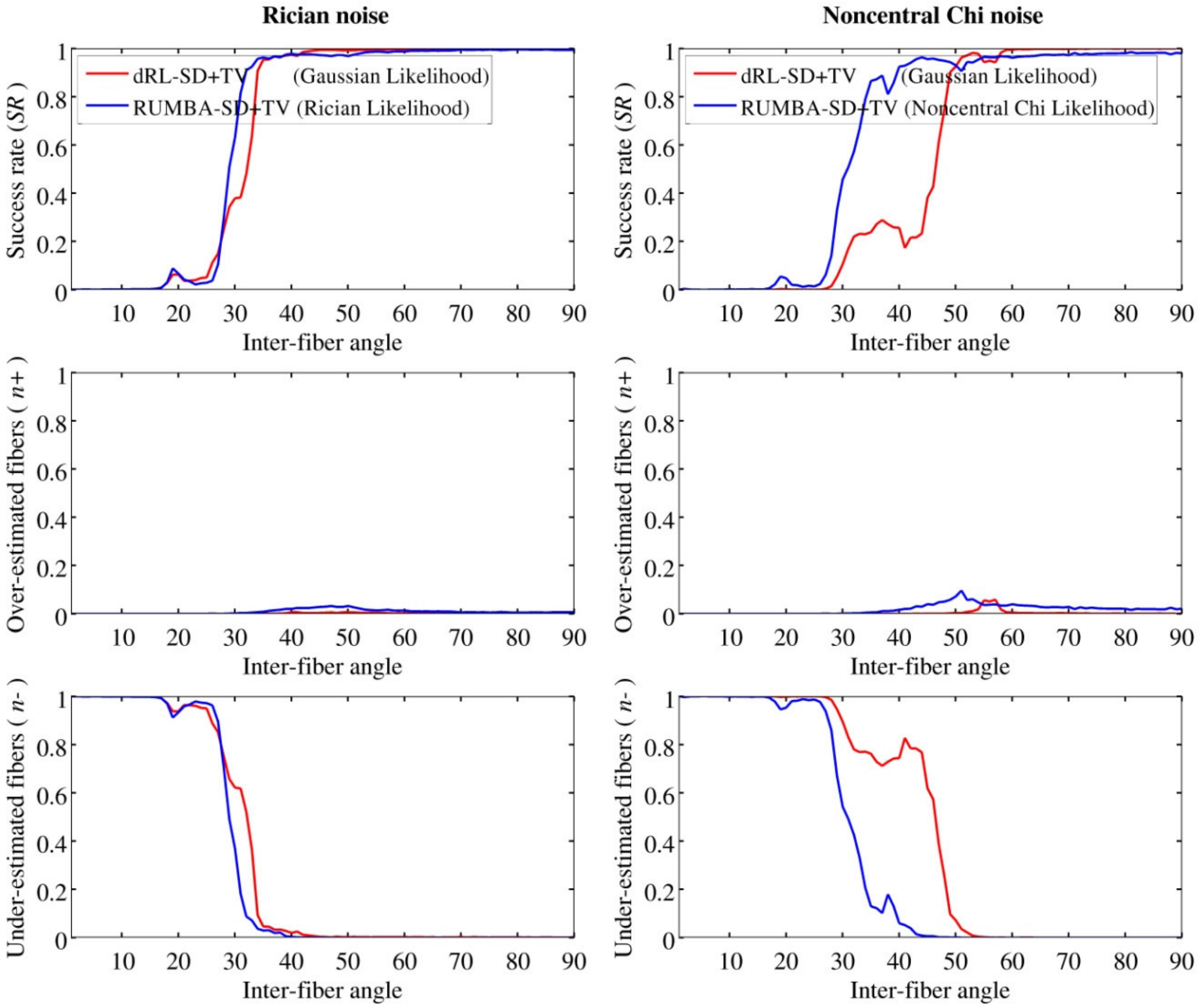

**Fig G**. Quantification of the reconstruction accuracy of RUMBA-SD+TV (blue color) and dRL-SD+TV (red color) in terms of the success rate (SR) and the mean number of over-estimated ($n^+$) and under-estimated ($n^-$) fiber populations, as a function of the inter-fiber angle in the 90 synthetic phantoms. The continuous lines in each plot represent the mean values for each method. This analysis refers to results using a dictionary created with the same diffusivities utilized to generate the data and a level of noise with a SNR = 15.

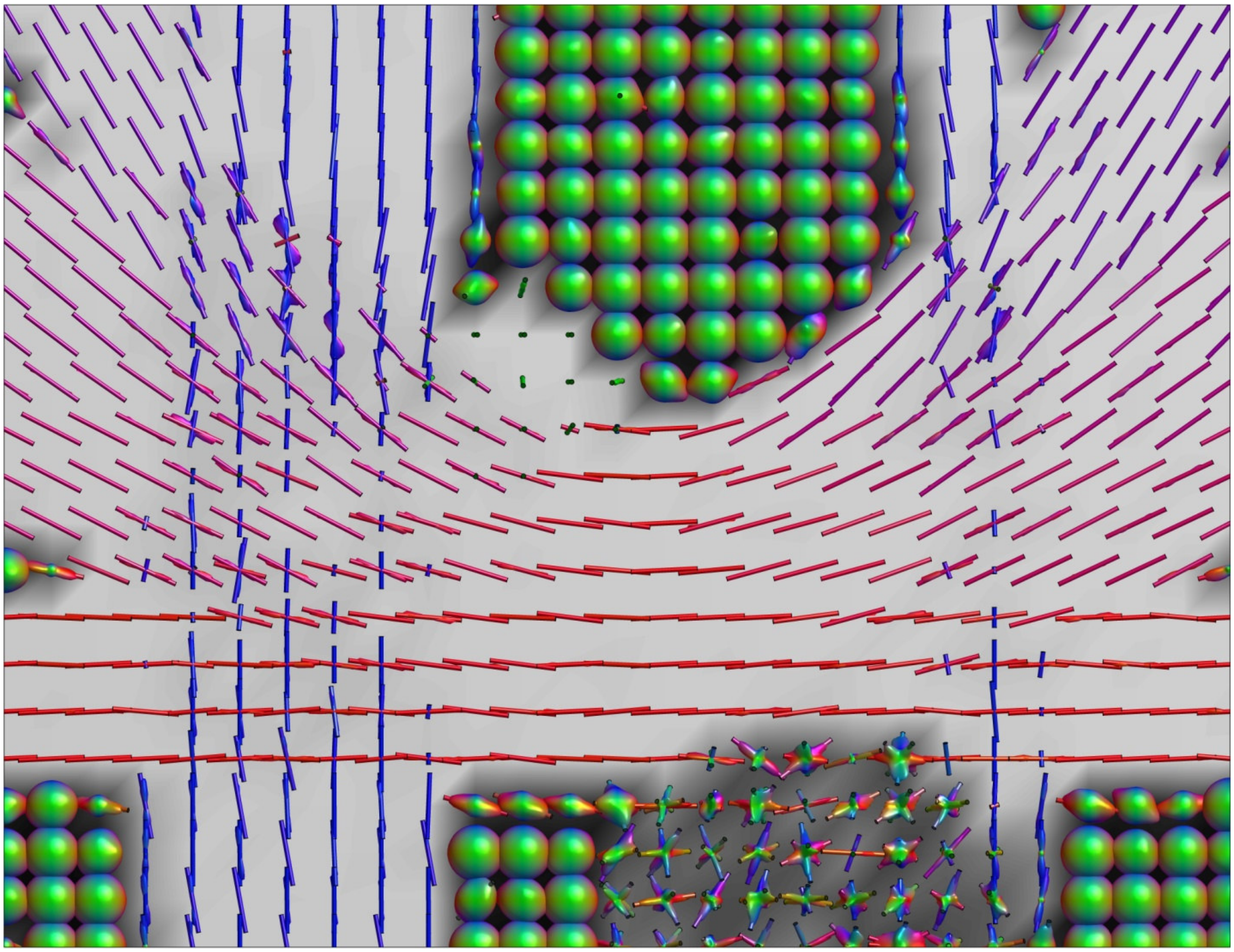

**Fig H**. Visualization of the fiber ODFs and their peaks (plotted as thin cylinders) reconstructed from the SMF-based data generated with SNR = 20 in a coronal slice of the “HARDI Reconstruction Challenge 2013” phantom. Depicted fiber ODF profiles correspond to the estimates from **dRL-SD** using 400 iterations.

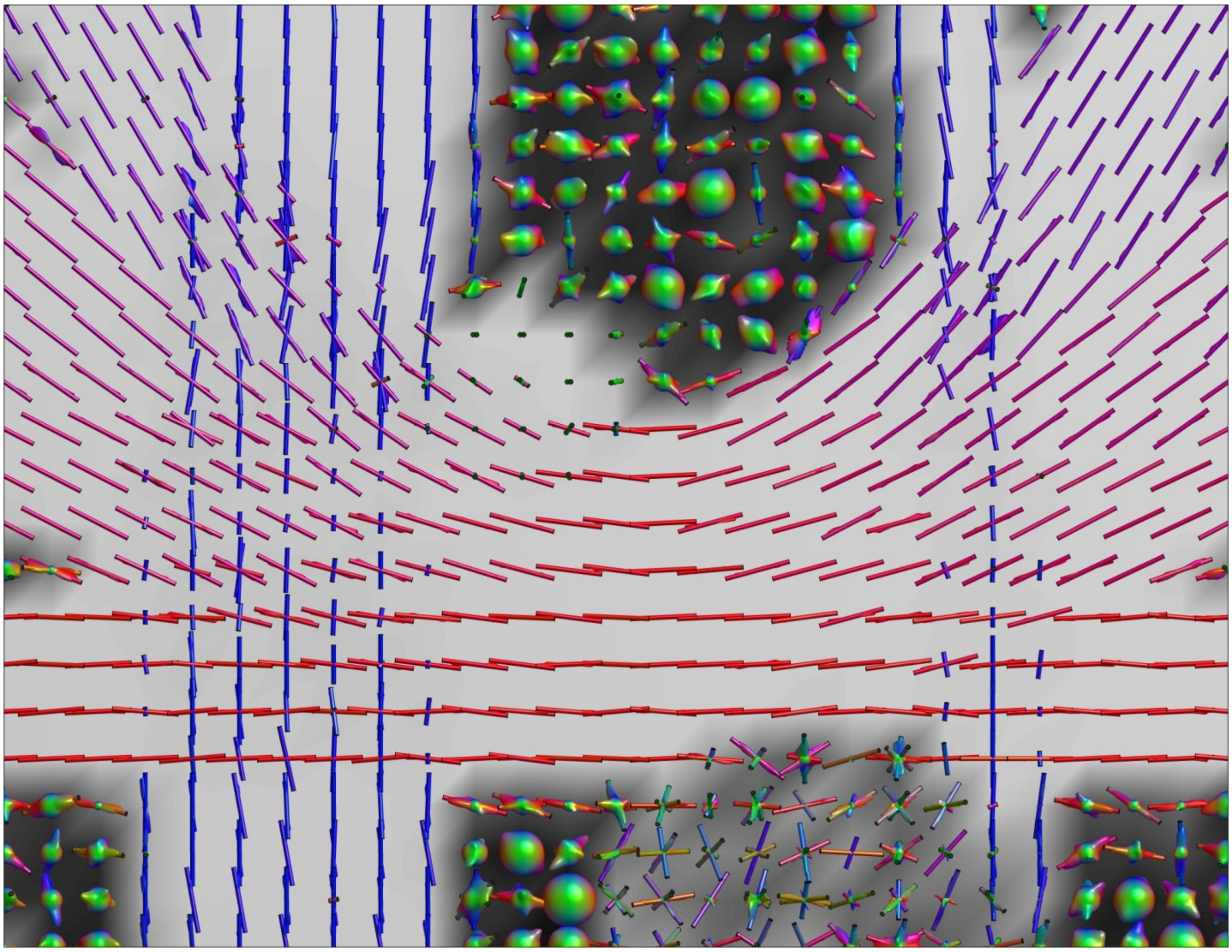

**Fig I**. Visualization of the fiber ODFs and their peaks (plotted as thin cylinders) reconstructed from the SMF-based data generated with SNR = 20 in a coronal slice of the “HARDI Reconstruction Challenge 2013” phantom. Depicted fiber ODF profiles correspond to the estimates from **RUMBA-SD** using 400 iterations.

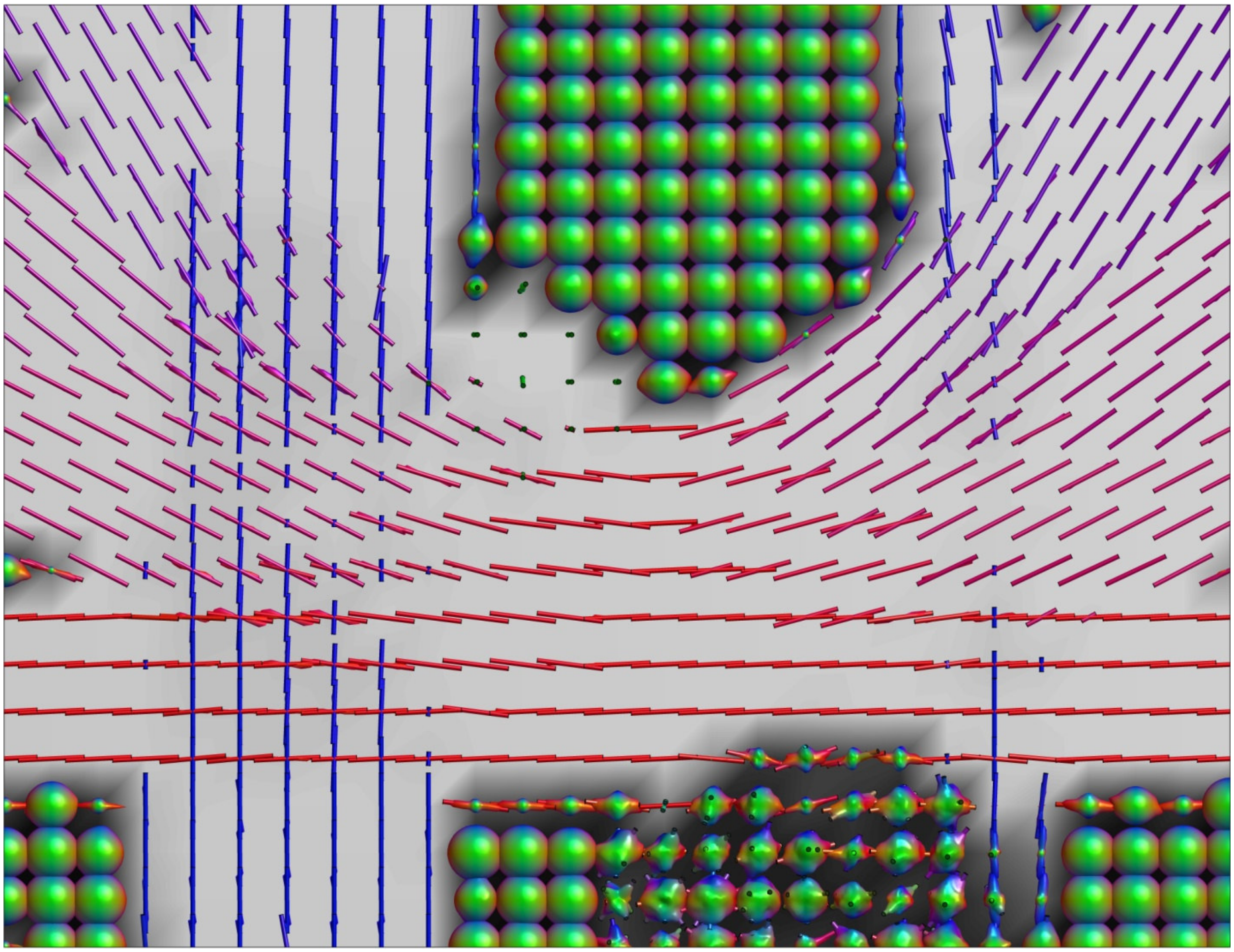

**Fig J**. Visualization of the fiber ODFs and their peaks (plotted as thin cylinders) reconstructed from the SMF-based data generated with SNR = 20 in a coronal slice of the "HARDI Reconstruction Challenge 2013" phantom. Depicted fiber ODF profiles correspond to the estimates from **dRL-SD+TV** using 400 iterations.

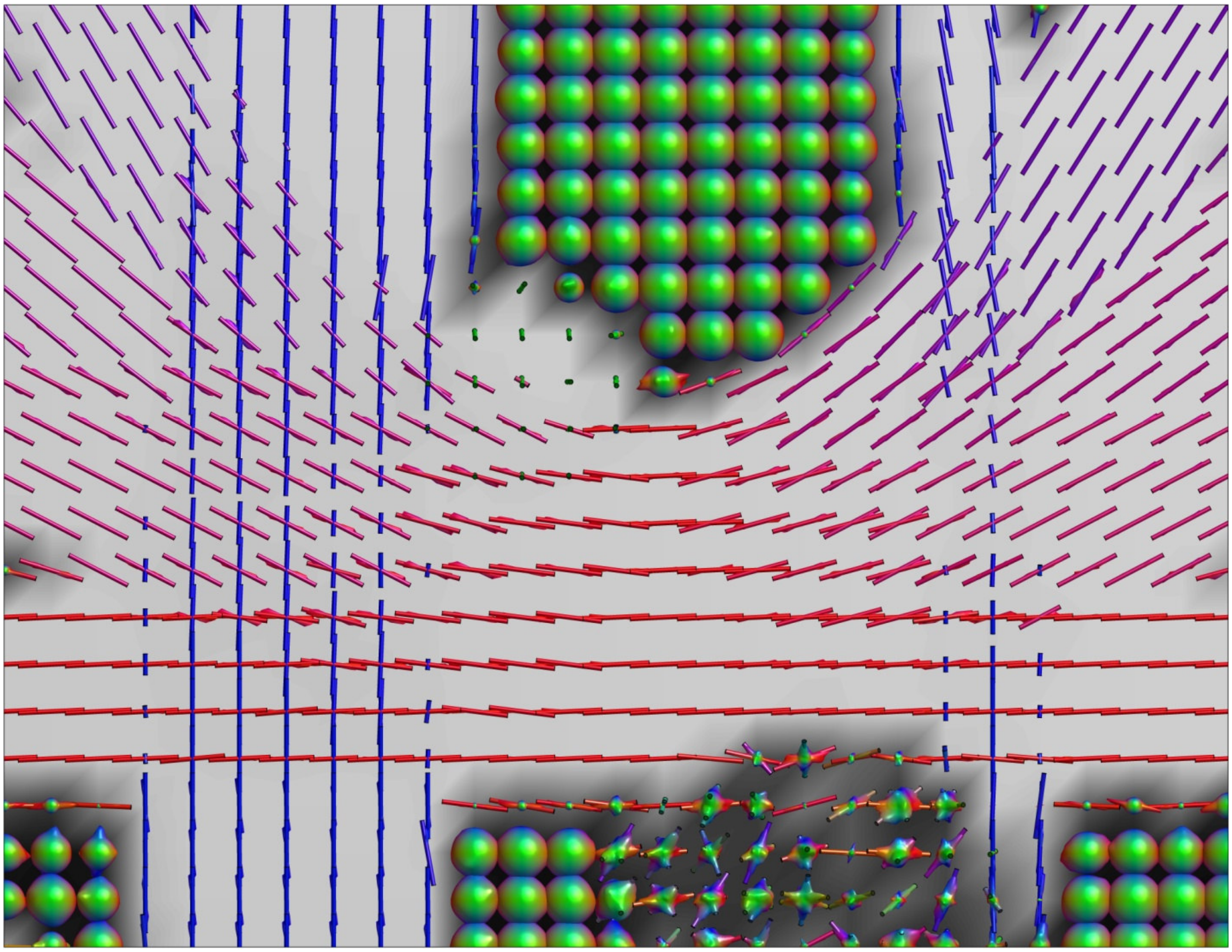

**Fig K**. Visualization of the fiber ODFs and their peaks (plotted as thin cylinders) reconstructed from the SMF-based data generated with SNR = 20 in a coronal slice of the "HARDI Reconstruction Challenge 2013" phantom. Depicted fiber ODF profiles correspond to the estimates from **RUMBA-SD+TV** using 400 iterations.

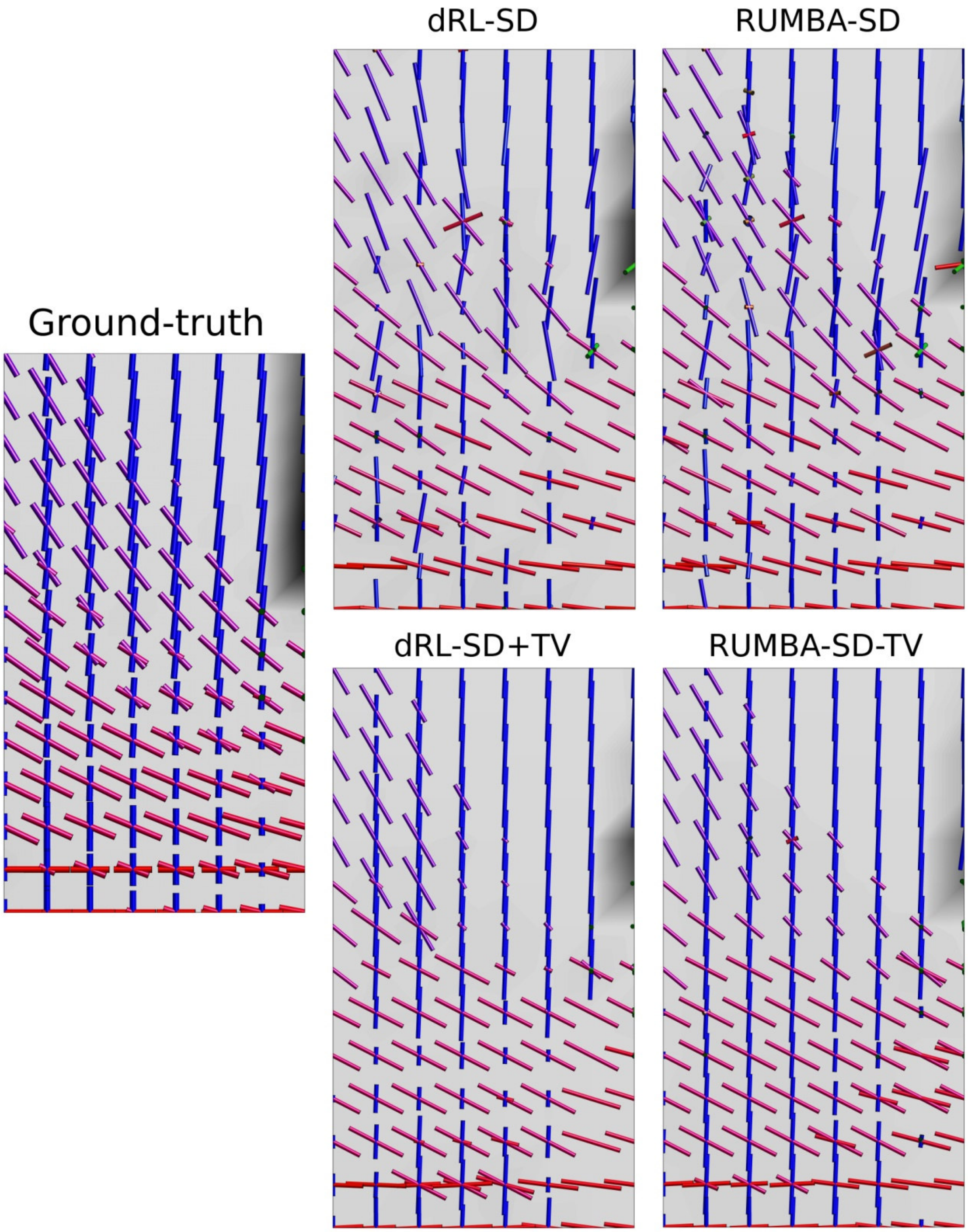

**Fig L**. Main peaks from the fiber ODFs estimated in the “HARDI Reconstruction Challenge 2013” phantom. Visualization of the main peaks extracted from the fiber ODFs reconstructed

from the SMF-based data generated with SNR = 20 in a complex region of the "HARDI Reconstruction Challenge 2013" phantom. Results are based on reconstructions using 1000 iterations. Peaks are visualized as thin cylinders.

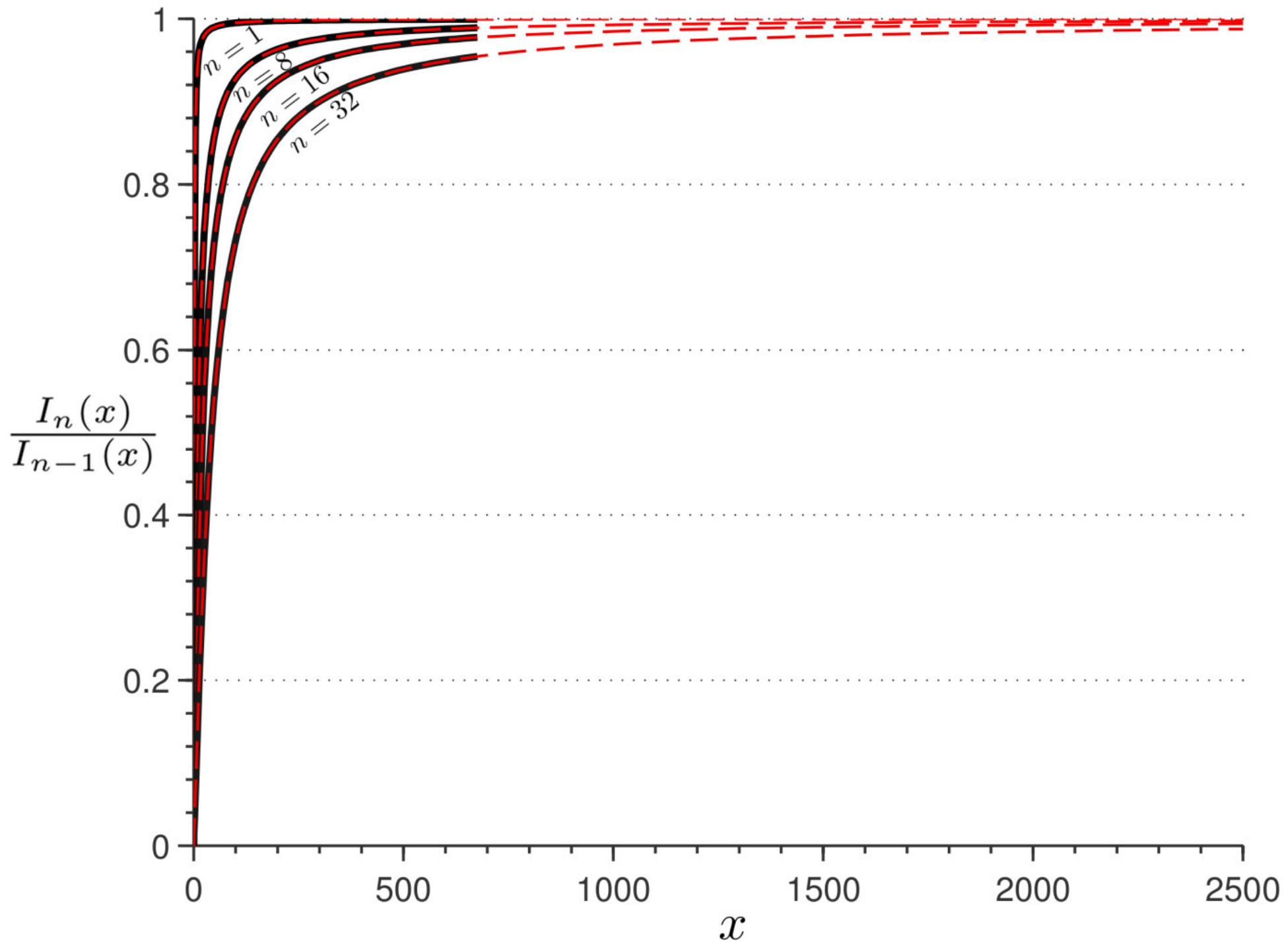

**Fig M**. Ratio of modified Bessel functions of first kind. Black continuous curves denote the exact values computed by means of the evaluation of the ratio of the individual Bessel functions. The fast divergence towards infinity of the individual functions does not allow evaluating this expression for the whole range of values. Red discontinuous curves denote the values computed by means of the Perron continued fraction approximation in Appendix C in S1 File, in the whole range of values.